\documentclass[transmag,10pt,final]{IEEEtran}

\usepackage{silence}
\usepackage{cite}

\usepackage{amsmath,amssymb,amsthm,fixmath}

\usepackage{xcolor}

\usepackage[inline]{enumitem}

\usepackage{siunitx}
\DeclareSIUnit{\dBm}{dBm}

\usepackage{caption}
\usepackage{graphicx}
\usepackage{subcaption}

\usepackage{epstopdf}
\usepackage{cuted}
\usepackage{multirow,booktabs}
\usepackage{diagbox}

\usepackage{optidef}

\usepackage[hidelinks]{hyperref}

\usepackage[shortcuts,acronym,automake]{glossaries}
\makeglossaries
\newacronym{3D}{3D}{three-dimensional}
\newacronym{5G}{5G}{Fifth Generation}
\newacronym{6G}{6G}{Sixth Generation}
\newacronym{ACK}{ACK}{acknowledgement}
\newacronym{AF}{AF}{amplify-and-forward}
\newacronym{ARQ}{ARQ}{automatic repeat request}
\newacronym{AWGN}{AWGN}{additive white Gaussian noise}
\newacronym{B5G}{B5G}{Beyond-Fifth-Generation}
\newacronym{BS}{BS}{base station}
\newacronym{CC}{CC}{chase combining}
\newacronym{CLARQ}{CLARQ}{Closed-Loop ARQ}
\newacronym{CQI}{CQI}{channel quality indicator}
\newacronym{CSI}{CSI}{channel state information}
\newacronym{O-CU}{O-CU}{open centralized unit}
\newacronym{DF}{DF}{decode-and-forward}
\newacronym{DL}{DL}{downlink}
\newacronym{O-DU}{O-DU}{open distributed unit}
\newacronym{DMH-HARQ}{DMH-HARQ}{Dynamic Multi-Hop HARQ}
\newacronym{DP}{DP}{dynamic programming}
\newacronym{E2E}{E2E}{end-to-end}
\newacronym{FBL}{FBL}{finite blocklength}
\newacronym{HAP}{HAP}{high altitude platform}
\newacronym{HARQ}{HARQ}{hybrid ARQ}
\newacronym{IBL}{IBL}{infinite blocklength}
\newacronym{IIoT}{IIoT}{industrial Internet-of-Things}
\newacronym{IR}{IR}{incremental redundancy}
\newacronym{LBT}{LBT}{Listen-Before-Talk}
\newacronym{LoS}{LoS}{line-of-sight}
\newacronym{LUT}{LUT}{look-up table}
\newacronym{mmWave}{mmWave}{millimeter-wave}
\newacronym{MIMO}{MIMO}{multi-input multi-output}
\newacronym{MDP}{MDP}{Markov decision process}
\newacronym{MEC}{MEC}{multi-access edge computing}
\newacronym{NACK}{NACK}{negative acknowledgement}
\newacronym{NOMA}{NOMA}{non-orthogonal multiple access}
\newacronym{NR}{NR}{New Radio}
\newacronym{NTN}{NTN}{non-terrestrial network}
\newacronym{OFDMA}{OFDMA}{orthogonal frequency-division multiple access}
\newacronym{O-RAN}{O-RAN}{Open-Radio Access Network}
\newacronym{RAN}{RAN}{Radio Access Network}
\newacronym{PER}{PER}{packet error rate}
\newacronym{PDF}{PDF}{probability density function}
\newacronym{RL}{RL}{reinforcement learning}
\newacronym{O-RU}{O-RU}{open radio hardware unit}
\newacronym{SNR}{SNR}{signal-to-noise ratio}
\newacronym{SIC}{SIC}{successive interference cancellation}
\newacronym{SINR}{SINR}{signal-to-interference-and-noise ratio}
\newacronym{TDMA}{TDMA}{time-division multiple access}
\newacronym{THz}{THz}{terahertz}
\newacronym{TWT}{TWT}{target wake time}
\newacronym{V2X}{V2X}{vehicule-to-everything}
\newacronym{VNF}{VNF}{virtualized network function}
\newacronym{vRAN}{vRAN}{virtualized radio access networks}
\newacronym{UAV}{UAV}{unmanned aerial vehicle}
\newacronym{UE}{UE}{user equipment}
\newacronym{UL}{UL}{uplink}
\newacronym{URLLC}{URLLC}{ultra-reliable low-latency communications}
\newacronym{SMO}{SMO}{Service Management and Orchestration}

\usepackage{tcolorbox}
\tcbuselibrary{many}
\newtheorem{theorem}{Theorem}% from 'amsthm'
\newtheorem{lemma}{Lemma}% from 'amsthm'
\usepackage[norelsize,linesnumbered,ruled]{algorithm2e}
\SetKwRepeat{Do}{do}{while}
\makeatletter
\newcommand{\removelatexerror} {\let\@latex@error\@gobble}
\makeatother

\newcommand{\superscript}[1]{^{\mathrm{#1}}}
\newcommand{\subscript}[1]{_{\mathrm{#1}}}

\newcommand{\diff}{\text{d}}

\newcommand{\highlightcolor}{blue}
\renewcommand{\highlightcolor}{black}

\newcommand{\revise}[1]{\highlight{#1}}
\renewcommand{\revise}[1]{#1}

\newcommand{\revisebox}[1]{\fcolorbox{\highlightcolor}{white}{#1}}
\renewcommand{\revisebox}[1]{#1}

\usepackage[textsize=tiny,colorinlistoftodos]{todonotes}
\makeatletter
\define@key{todonotes}{bh}[]{% Comment by 
	\setkeys{todonotes}{author=\textbf{Bin}, color=lime!30}}%
\define@key{todonotes}{vs}[]{% Comment by 
	\setkeys{todonotes}{author=\textbf{Vincenzo}, color=blue!30}}%
\define@key{todonotes}{asif}[]{% Comment by 
		\setkeys{todonotes}{author=\textbf{Asif}, color=teal!30}}%
\makeatother

\begin{document}

%\title{Reliable O-RAN-Inspired Latency-Constrained\\Wireless Transport Network in 6G}
\title{DMH-HARQ: Reliable and Open Latency-Constrained Wireless Transport Network}

%% Censor authors and affliations for the double-blind review
%\author{
%	\IEEEauthorblockN{%Anonymous authors
%		Author~1\IEEEauthorrefmark{1}, Author~2\IEEEauthorrefmark{2}, Author~3\IEEEauthorrefmark{3}, Author~4\IEEEauthorrefmark{3}, Author~5\IEEEauthorrefmark{3}, Author~6\IEEEauthorrefmark{2}, and Author~7\IEEEauthorrefmark{1}\IEEEauthorrefmark{4}
%}
%	\IEEEauthorblockA{
%		%Anonymous affiliations
%		Affiliation~1\IEEEauthorrefmark{1},
%		Affiliation~2\IEEEauthorrefmark{2},
%		Affiliation~3\IEEEauthorrefmark{3},
%		Affiliation~4\IEEEauthorrefmark{4}
%	}%
%}

 % \author{Bin~Han,
	% 	Muxia~Sun,
	% 	Yao~Zhu,
	% 	Vincenzo~Sciancalepore,
 %        Mohammad~Asif~Habibi,\\
 % 		Yulin~Hu,
	% 	Anke~Schmeink,
	% 	 Yan-Fu~Li, and
	%  	Hans~D.~Schotten
	%  	\thanks{B. Han, M. A. Habibi and H. D. Schotten are with Rheinland-Pf\"alzische Teschnische Universit\"at Kaiserslautern-Landau (RPTU). M. Sun and Y.-F. Li are with Tsinghua University. Y. Zhu, Y. Hu, and A. Schmeink are with RWTH Aachen University. V. Sciancalepore is with NEC Laboratories Europe. Y. Zhu and Y. Hu are with Wuhan University. H. D. Schotten is with the German Research Center for Artificial Intelligence (DFKI). B. Han (bin.han@rptu.de) is the corresponding author.}
 % }

\author{Bin Han,~\IEEEmembership{Senior Member,~IEEE,}
Muxia Sun,~\IEEEmembership{Member,~IEEE,}
Yao Zhu,~\IEEEmembership{Member,~IEEE,}\\
Vincenzo Sciancalepore,~\IEEEmembership{Senior Member,~IEEE,}
Mohammad Asif Habibi,
Yulin Hu,~\IEEEmembership{Senior Member,~IEEE,}\\
Anke Schmeink,~\IEEEmembership{Senior Member,~IEEE,}
Yanfu Li,~\IEEEmembership{Senior Member,~IEEE,}
Hans D. Schotten,~\IEEEmembership{Member,~IEEE}
        % <-this % stops a space
\thanks{B. Han, M. A. Habibi, and H. D. Schotten are with RPTU Kaiserslautern-Landau. M. Sun and Y.-F. Li are with Tsinghua University. Y. Zhu, Y. Hu, and A. Schmeink are with RWTH Aachen University. V. Sciancalepore is with NEC Laboratories Europe.  Y. Zhu and Y. Hu are with Wuhan University.  H. D. Schotten is with the German Research Center for Artificial Intelligence (DFKI). B. Han (bin.han@rptu.de) and M. Sun (muxiasun@mail.tsinghua.edu.cn) are the corresponding authors.}% <-this % stops a space
\thanks{The work of B. Han, M. A. Habibi, A. Schmeink, and H. D. Schotten was supported in part by the German Federal Ministry of Education and Research in the programme of ``Souver\"an. Digital. Vernetzt.'' joint projects 6G-RIC (16KISK028), Open6GHub (16KISK003K/16KISK004/16KISK012), and Open6GHub+ (16KIS2406). The work of Y. Zhu and Y. Hu was supported in part by the National Key R\&D Program of China under Grant 2023YFE0206600, NSFC Grant 62471341, and by the Hubei Provincial Science and Technology Cooperation Project under Grant 2025EHA040.  The work of V. Sciancalepore was partially supported by SNS JU Project 6G-GOALS (GA no. 101139232).}
}

\markboth{IEEE/ACM Transactions on Networking}%
{Han \MakeLowercase{\textit{et al.}}: DMH-HARQ: Towards Latency-Constrained Wireless Transport Networks in 6G Open RAN}

%\IEEEpubid{0000--0000/00\$00.00~\copyright~2024 IEEE}

\maketitle


\maketitle

\begin{abstract}
	
The extreme requirements for high reliability and low latency in the upcoming \ac{6G} wireless networks are challenging the design of multi-hop wireless transport networks. Inspired by the advent of the virtualization concept in the wireless networks design and \emph{openness} paradigm as fostered by the \ac{O-RAN} Alliance, we target a revolutionary resource allocation scheme to improve the overall transmission efficiency.

In this paper, we investigate the problem of {\ac{ARQ} in} multi-hop \ac{DF} relaying in the \ac{FBL} regime, and propose {a dynamic scheme of multi-hop \ac{HARQ}}, which maximizes the \ac{E2E} communication reliability in the wireless transport network. We also propose an integer \ac{DP} algorithm to efficiently solve the optimal \ac{DMH-HARQ} strategy. Constrained within a certain time frame to accomplish E2E transmission, our proposed approach is proven to outperform {the conventional listening-based cooperative \ACRshort{ARQ}, as well as any static \ACRshort{HARQ} strategy, regarding the \ac{E2E} reliability. It is applicable without dependence on special delay constraint, and is particularly competitive for long-distance transport network with many hops.}
\end{abstract}


\IEEEpeerreviewmaketitle

\glsresetall

\section{Introduction}\label{sec:introduction}

\revise{
\Ac{URLLC} is pushing wireless transport beyond fiber-based backhaul toward multi-hop wireless chains that stitch together terrestrial small cells and non-terrestrial segments within virtualized, \ac{O-RAN}~\cite{Samsung_vran2019,ORAN_GC2021,nuberu_2021}. In such deployments, fronthaul, midhaul, and backhaul flows may traverse sequences of \ac{O-RU}, \ac{O-DU}, and \ac{O-CU} nodes hosted on O-Cloud and orchestrated by the SMO, with the specific chain length and composition dictated by functional splits and dynamic placement. 

The shift to \ac{mmWave} and (sub-)THz for capacity and positioning tightens line-of-sight constraints and shrinks cell footprints, making dense, relay-rich topologies economically preferable to ubiquitous fiber~\cite{TO2022wireless,JCS+2022wireless}. At the same time, 3D coverage objectives fold in \ac{NTN} elements—\acp{UAV}, \acp{HAP}, and satellites—whose heterogeneous channels and feedback latencies further complicate transport~\cite{JHH+2021road}. While relays have long been used to extend coverage and combat fading~\cite{athanasopoulos2010multi,karmakar2011diversity}, URLLC alters the operating point: packets are short (finite blocklength), feedback and control overheads are non-negligible, and strict end-to-end (E2E) deadlines mean that each retransmission on one hop directly consumes the time/frequency budget of successor hops. Under these conditions, classical relaying and ARQ/\ac{HARQ} designs (tuned for long codes and per-hop optimization) systematically misestimate reliability and cannot coordinate retransmissions and forwarding at the granularity and speed demanded by URLLC~\cite{Hu2016CapacityRelay,makki2014finite}.

{\it Challenges.} The first challenge is reliability composition under \ac{FBL}: with short codes, per-hop decoding errors remain non-negligible and compound across hops, making E2E reliability exquisitely sensitive to how retransmissions and forwarding are budgeted within a tight delay. Second, resources are coupled across hops; allocating more symbols to a retransmission on one hop reduces the symbols—and thus reliability—available to successor hops. Third, links at mmWave and (sub-)THz under strict power limits and interference are brittle, while \ac{NTN} segments (\acp{UAV}, \acp{HAP}, satellites) introduce heterogeneous channels and feedback delays. Fourth, open and disaggregated \ac{O-RAN} deployments with flexible functional splits (\ac{O-RU}/\ac{O-DU}/\ac{O-CU}, O-Cloud, SMO) yield relay chains whose length and per-hop characteristics vary with placement and load, challenging fixed frame designs and static schedulers.}

\begin{figure}
	\centering
	\revisebox{\includegraphics[height=2.3 in, width=3.5in]{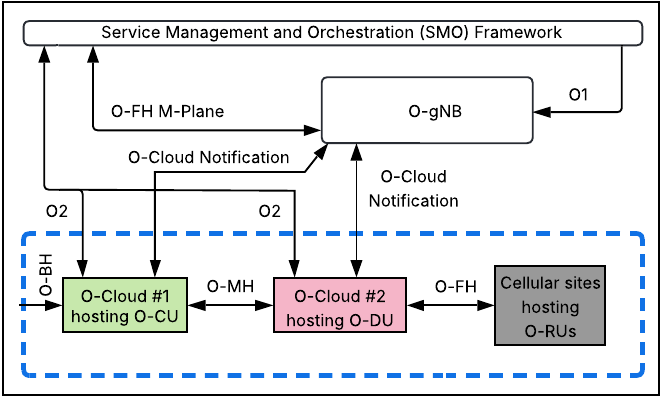}}
	\captionsetup{font={color=\highlightcolor}}
	\caption{\revise{A conceptual depiction of the O-RAN architecture is presented. This article concentrates on the areas outlined by the dashed blue box, where the proposed transmission scheme enhances E2E communication reliability.}}
	\label{fig:architecture}
\end{figure}

\revise{
{\it Limitations of existing methods.} Classical relaying assumes long codes and asymptotic coding gains, which do not hold in the \ac{FBL} regime of short packets~\cite{Hu2016CapacityRelay,makki2014finite}. Standard ARQ/\ac{HARQ} mechanisms are tuned to per-hop reliability or subframe-level allocation and do not perform outcome-conditioned reallocation under a global E2E deadline, thereby neglecting the inter-hop coupling central to URLLC. Fixed retransmission counts, timers, or frame partitions cannot adapt to instantaneous decode outcomes, hop heterogeneity, or \ac{O-RAN}-driven topology changes. Even listening-based cooperative \ac{ARQ}~\cite{goel2021listen} typically moves resources at the frame/subframe timescale: when packets are short, symbol-level decisions become pivotal yet are unsupported.

{\it Contributions.} We propose a dynamic \ac{HARQ} scheme for multi-hop relaying in the \ac{FBL} regime that reallocates—at \emph{channel-use (symbol) granularity}—the remaining radio budget between the current hop's repeats and successor hops, conditioned on instantaneous decode outcomes. The design is \ac{O-RAN}-compatible (Fig.~\ref{fig:architecture}) and targets open, disaggregated transport. Our key insight is to treat retransmissions and forwarding as a single coupled E2E optimization rather than per-hop tuning, updating the residual budget after each (re)transmission using posterior success probabilities under \ac{FBL}. We formalize the scheduler and provide a dynamic programming (DP) algorithm that optimizes the sequence of (re)transmissions across hops under latency and reliability targets. 

In summary, our contributions are as follows: $i$) a dynamic, symbol-level \ac{HARQ} framework that explicitly couples retransmission and forwarding under a strict E2E deadline, $ii$) an \ac{FBL}-based analysis that quantifies reliability composition across hops, and $iii$) a DP-based optimal scheduler with implementable policies aligned with \ac{O-RAN} disaggregation. Compared to static \ac{HARQ}, subframe-level listening-based cooperative \ac{ARQ}~\cite{goel2021listen}, and long-code-inspired relaying~\cite{athanasopoulos2010multi,karmakar2011diversity}, our approach delivers improved reliability–latency trade-offs for \ac{URLLC} over \ac{O-RAN}-inspired wireless transport.
}

The remainder of the paper is organized as follows: {We begin with Section~\ref{sec:related} to provide a brief review to selected literature in related field. 
	Section~\ref{sec:setup} follows to describe and formulate} the investigated problem, which is then analyzed in Section~\ref{sec:analyses} and solved in Section~\ref{sec:design}. The proposed methods are numerically evaluated in Section~\ref{sec:simulation}, followed by some further discussions in Section~\ref{sec:discussion}. To the end, we {close the paper with our conclusion and out  s in Section~\ref{sec:conclusion}.}

{
\section{Related Work}\label{sec:related}

The problem of multi-hop relay has been attracting great interest in the wireless field since long, for it has a rich potential to offer extended cell coverage and improved channel capacity~\cite{athanasopoulos2010multi}, {particularly in the \ac{O-RAN} architecture \cite{WG7CellArchReq001, electronics10172162}}. Generally, there are two families of relay technologies, namely \ac{DF} and \ac{AF}. Albeit the disadvantage of higher implementation complexity, \ac{DF} comprehensively outperforms \ac{AF} in latency-reliability performance~\cite{chaaban2016multi}, and is therefore preferred by many researchers. Literature has investigated multi-hop \ac{DF} from various perspectives, including modulation scheme~\cite{dhaka2012performance}, selective relaying route~\cite{farhadi2010fixed}, and relay station placement~\cite{ibrahim2016optimization}, and \ac{ARQ} strategy~\cite{maagh2010dynamic}.

Over the past few years, the emerging wireless use scenario of \ac{URLLC} {in \ac{O-RAN}} is challenging off-the-shelf multi-hop relay. The extreme requirement of low \ac{E2E} latency in {\ac{O-RAN}} \ac{URLLC} is leading to even more stringent time constraint over each individual hop {\cite{9750106, 9838575}}, which is barely achievable with conventional technologies.
%\todo{
	%%\begin{itemize}
	%Multi-hop DF SotA:\begin{enumerate}
		%	\item fixed maximal HARQ attempts;
		%	\item listening-based dynamic HARQ. 
		%	\end{enumerate}
	%	Generally fixed codec.
	%	
	%	Multi-hop URLLC:\begin{itemize}
		%	\item \href{https://ieeexplore.ieee.org/abstract/document/8896993}{ref 1}
		%	\item \href{https://ieeexplore.ieee.org/abstract/document/9625389}{ref 2}
		%	\item \href{https://ieeexplore.ieee.org/abstract/document/9448948}{ref 3}
		%	\item \href{https://ieeexplore.ieee.org/abstract/document/9685765}{ref 4}
		%	\item \href{https://ieeexplore.ieee.org/abstract/document/9186171}{ref 5}
		%	\end{itemize}}
To meet such a stringent latency requirement, \ac{URLLC} transmissions are likely carried out with short packet, i.e., the codewords cannot considered as infinite long. With such so-called \ac{FBL} codes, many theories and propositions in classical information theory fail to apply since the Shannon limit is no more asymptotically achievable. To accurately characterize the relationship between achievable data rate and reliability in the \ac{FBL} regime, \emph{Polyanskiy} et al. derived in~\cite{polyanskiy2010channel} a closed-form expression for the single-hop transmission in \ac{AWGN} channels. Later on, this \ac{FBL} characterization was generalized, extended into Gilbert-Elliot channels~\cite{Polyanskiy2009gilbert} and quasi-static flat-fading channels~\cite{quasi_static}. Recently, the \ac{FBL} model in random access channel has also been analyzed~\cite{Effros2018random}. On the basis of those results, the performance \ac{FBL} communication systems has been investigated in context of various wireless networking technologies, such as \ac{MIMO}~\cite{Zhang2021MIMO}, \ac{OFDMA}\cite{han2021fairness}, \ac{NOMA}~\cite{Sun2018NOMA} and \ac{MEC}~\cite{Zhu2022MEC}.

Specifically for \ac{FBL} relay networks, the system performance are generally investigated in a deterministic manner in the most existing  works~\cite{Hu2016CapacityRelay,Hu2018CSIRelay,Pan2019UAVRelay,Hu2019SWIPTRelay,Hu2020NOMARelay}. In particular, the authors of~\cite{Hu2016CapacityRelay} studied the performance bound of two-hop \ac{FBL} relay system with perfect \ac{CSI}, which was later extended to the scenarios with only partial \ac{CSI} available~\cite{Hu2018CSIRelay}. Moreover, the authors of~\cite{Pan2019UAVRelay} investigated a potential use case where \ac{UAV} behaves as the relay between source and destination to improve the system performance. Energy harvesting was introduced in~\cite{Hu2019SWIPTRelay} where the relay solely relies on the harvested energy to forward the information. The authors of~\cite{Hu2020NOMARelay} studied the impact of imperfect \ac{SIC} in \ac{NOMA} schemes on relay system that causes by the FBL transmission error.

Another interesting  \ac{FBL} use case is the closed-loop communication, which is not only common in \ac{MEC} {scenarios} but also strongly related to \ac{FBL} relay networks, because it behaves in a similar way as a two-hop relay chain does. In a previous work of ours we proposed for this use case the so-called \ac{CLARQ} algorithm~\cite{han2022clarq}, which dynamically re-allocates blocklengths between uplink retransmission and downlink reception, so as to minimize the closed-loop error rate. Moreover, in \cite{han2021time} we analyzed the optimal one-shot transmission scheme, which is a special case of \ac{CLARQ}, under both constraints of latency and energy consumption.
%\todo{introduce the clarq: ``\dots  the dynamic \ac{ARQ} mechanism that we proposed in a previous study~\cite{han2022clarq} for reliable closed-loop message exchange, which is a special case of two-hop transmission chain, namely \ac{CLARQ}\dots''}
%
%Discarding the data processing that is executed on MEC server and independent from the radio transmission, we can model the closed-loop communication that is addressed by the CL\ac{ARQ} protocol equivalent to a two-hop \ac{DF} relaying chain, where the uplink transmission is the first hop and downlink the second. The second hop transmission only makes sense if the first hop transmission has succeeded. By adding more hops into this chain, this model can be generalized into an $I$-hop version.
%\end{itemize}
}

\section{Problem Setup}\label{sec:setup}
In context of the \ac{O-RAN} architecture~\cite{nuberu_2021}, we consider a multi-hop data link in the wireless transport network, which has three stages: the \ac{O-RU} is connected to the \ac{O-DU} over a wireless fronthaul, the \ac{O-DU} to the \ac{O-CU} over a wireless midhaul, and the \ac{O-CU} to the core network over a wireless backhaul. Upon the specific function splitting and orchestration scheme, a VNF may be flexibly placed at the \ac{O-RU}, \ac{O-DU}, or \ac{O-CU}. In the latter two cases, data involved with the \ac{VNF} must be decoded and forwarded over one or multiple hops from the \ac{O-RU} to the implementation node of the \ac{VNF} in the \ac{UL}, or in the inverse direction for the \ac{DL}, or both for a closed-loop data link. While the wireless fronthaul has usually only one hop, both the midhaul and backhaul links can often have multiple hops, due to the high communication distance (typically between \SIrange{20}{40}{\kilo\meter} for the midhaul and \SIrange{20}{300}{\kilo\meter} for the backhaul~\cite{gomes2019reducing}). Thus, the wireless transport network can be modeled as a multi-hop \ac{DF} relaying chain, which is supposed to meet a stringent latency constraint.

\subsection{Multi-Hop DF and DMH-HARQ}\label{subsec:setup_scenario}
Now consider a multi-hop relaying chain with $I+1$ nodes over $I$ hops, each applying the \ac{DF} strategy based on simple {\ac{HARQ}} without  combining. We assume that all channels of the $I$ hops are i.i.d., undergoing Gaussian noise and Rician fading. We assume that every decoding process, together the associated feedback of \ac{ACK} or \ac{NACK} to the predecessor node for the \ac{HARQ} procedure,  makes the same delay of $T\subscript{l}$. The overall transmission to a maximal \ac{E2E} delay is limited to $T\subscript{max}$. 
%The feedback latency to transmit \ac{ACK} or \ac{NACK} for the \ac{HARQ} procedure is neglected without loss of generality. 
For simplification of the analysis, we consider for every node the same symbol rate $f\subscript{s}$, modulation order {$Q\subscript{m}$} (and therefore the bit-time length $T\subscript{b}$=$\frac{1}{{Q\subscript{m}}f\subscript{s}}$), transmission power $P$, and noise power $N$. The transmitting node of each hop is capable of flexibly adjust its channel coding rate in every (re-)transmission attempt independently from others. We consider in this study a block fading scenario, where the channel $h_i$ of each hop $i\in\mathcal{I}=\{1,2,\dots I\}$ remains consistent over the entire time frame of $T\subscript{max}$. Each transmitting node of hop $i$ possesses the perfect \ac{CSI} of $h_i$, while knowing only its position in the relaying chain and the statistical \ac{CSI} of all other hops.

\begin{figure*}[!htbp]
	\centering
	{\includegraphics[width=.9\linewidth]{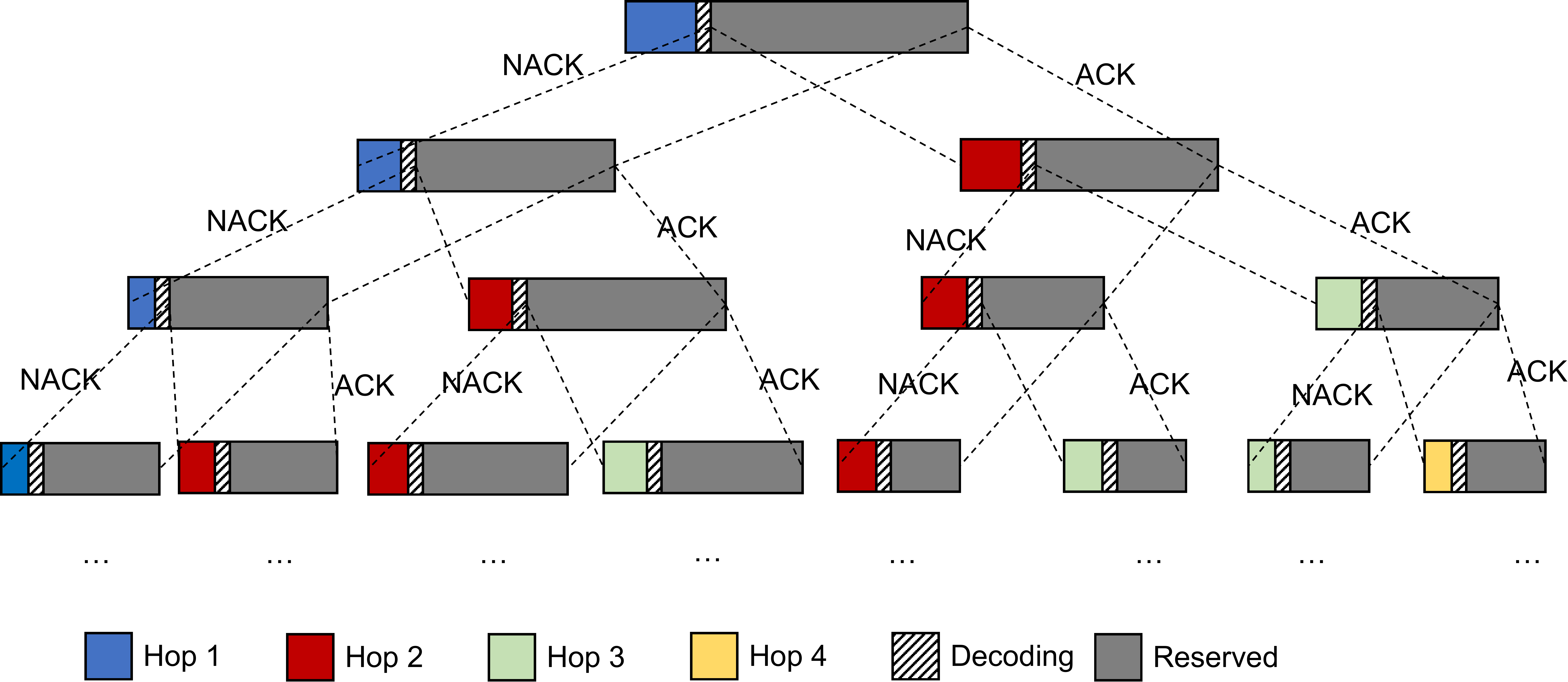}}%
	\caption{{Part of a {DMH-HARQ} schedule for 4-hop {DF} chain, beginning with the initial state $s_0=(T\subscript{max},4,0,0)$. At each (re)transmission attempt, an optimal blocklength is calculated for consumption, and a fixed time for decoding and feedback is required. Upon the decoding result, the remaining (reserved) blocklength is dynamically assigned regarding the decoding result: either for next hops upon success (ACK), or for retransmission on the same hop upon failure (NACK). The states $s_0$, $s_1$, and $s_2$ are illustrated with time frame details.}}
	\label{fig:dmh-harq}
\end{figure*}

Here we consider a problem of dynamic radio resource allocation among different \ac{HARQ} slots over the $I$-hop DF chain.
%adopt the principle of dynamic blocklength allocation in \ac{CLARQ} protocol, but in a generalized open-loop multi-hop DF relaying scenario. 
On every hop, the transmitting node can individually adjust its coding rate for the next (re)transmission. Since the modulation order and symbol rate are fixed, this will determine the time span of the transmission slot, which cannot exceed the remaining time of the entire transmission frame. If the message is successfully decoded, the successor node will take the turn and further forward the message, unless it is the sink node, i.e. the $(I+1)\superscript{th}$. If the decoding fails, the receiving node will trigger the \ac{HARQ} procedure by sending back a \ac{NACK} to the transmitting node, which then keeps re-transmitting the message until either a successful reception or a timeout, i.e. when the remaining time falls insufficient to support any further effective transmission. In every individual repeat, the node is able to readjust its coding rate. We consider a timer embedded in the header of every message, so that every node is aware of the time remaining from the $T\subscript{max}$ frame. We name this proposed approach as \ac{DMH-HARQ}, which is briefly illustrated in Fig.~\ref{fig:dmh-harq}{ in context of an example 4-hop DF chain}.

\subsection{Binary Schedule Tree}\label{subsec:binary_schedule_tree}
From Fig.~\ref{fig:dmh-harq} we see that the complete schedule of \ac{DMH-HARQ} has a structure of complete binary tree $\mathfrak{T}$, which can be recursively defined as a 3-tuple $\mathfrak{T}=(L(\mathfrak{T}), S(\mathfrak{T}), R(\mathfrak{T}))$, where $S(\mathfrak{T})$ is a singleton set containing the root node of $\mathfrak{T}$, $L(\mathfrak{T})$ and $R(\mathfrak{T})$ are the binary trees rooted on the left and right children of the root node, respectively:
\begin{align}
		&L( \mathfrak{T})
		=\begin{cases}
			\emptyset&S(\mathfrak{T})=\emptyset,\\
			\left(L(L(\mathfrak{T})), S(L(\mathfrak{T})), R(L(\mathfrak{T}))\right)&\text{o.w.},
		\end{cases}\label{eq:left_branch_binary_tree}\\
		&R( \mathfrak{T})
		=\begin{cases}
			\emptyset&S(\mathfrak{T})=\emptyset,\\
			\left(L(R(\mathfrak{T})), S(R(\mathfrak{T})), R(R(\mathfrak{T}))\right)&\text{o.w.}
		\end{cases}\label{eq:right_branch_binary_tree}
\end{align}

Each node $s$ in the tree representing a certain system state, which can be defined as a {4-}tuple $s=(t_s, i_s, k_s, \tau_s)$ where $t_s\in\mathbb{R}^+$ is the remaining time, $i_s\in\mathbb{N}$ is the remaining hops to transmit the message, $k_s\in\mathbb{N}$ is the number of failed \ac{HARQ} attempts on the current hop, and $\tau_s$ the accumulated time spent on these $k_s$ unsuccessful transmissions (excluding the feedback losses). {This system state concerns various aspects of the relaying session, involving the remaining resource, the transmissions to accomplish, and partial history of previous transmissions. Take the example in Fig.~\ref{fig:dmh-harq}, starting with the initial state $s_0=(T\subscript{max},4,0,0)$, where the schedule assigns $n_{s_0}$ bits for the first transmission, reserving therewith $r_{s_0}=T\subscript{max}-n_{s_0}T\subscript{b}-T\subscript{l}$ for future (re)transmissions. Upon the first transmission result, two possible succeding states can be $s_1=(r_{s_0},4,1,n_{s_0}T\subscript{b})$ and $s_2=(r_{s_0},3,0,0)$, respectively.} Now let $n_s\triangleq n(t_s, i_s, k_s, \tau_s)\in\mathbb{N}$ denote the the blocklength to be allocated {at state $s$}, according to the policy $n(\cdot)$, to the current \ac{HARQ} attempt on the current hop, i.e. the $(k_s+1)$\textsuperscript{th} \ac{HARQ} attempt on the {$(I-i_s+1)\superscript{th}$} hop, when there is $t_s$ time remaining. Thus, let $\mathfrak{T}_s$ denote a schedule tree with $S(\mathfrak{T}_s)=\{s\}$ and $\mathcal{S}$ the space of all feasible states, for all $s\in\mathcal{S}$, on the one hand we have for the left branch
\begin{equation}\label{eq:policy_tree_left_branch}
	\begin{split}
		&S(L(\mathfrak{T}_s))\\
		=&\begin{cases}
			\emptyset&t_s<T\subscript{min}^{(i_s)},\\
			\emptyset&i_s=0,\\
			\left\{(t_s-n_sT\subscript{b}-T\subscript{l}, i_s, k_s+1, \tau_s+n_sT\subscript{b})\right\}&\text{o.w.}
		\end{cases}
	\end{split}
\end{equation}
Here, $T\subscript{min}^{(i)}=\sum\limits_{j=1}^i\left(n_{\mathrm{min},j}T\subscript{b}+T\subscript{l}\right)$ where $n_{\mathrm{min},j}$ is the minimal blocklength of codeword on the $j$\textsuperscript{th} last hop (which is a function of the \ac{SNR} $\gamma_i$ over the $j$\textsuperscript{th} last hop). Therefore, in the first case of Eq.~\eqref{eq:policy_tree_left_branch}, $t_s<T\subscript{min}^{(i)}$ denotes the condition that a remaining time $t_s$ is insufficient to forward the message over the last $i$ hops even when no error occurs (i.e. with only one transmission attempt on each hop), under which the \ac{DMH-HARQ} procedure is terminated without success. The second case a successful completion of the DMH-HARQ over the entire chain (as there is $i=0$ hop remaining); the third case stands for the consequence when the current \Ac{HARQ} attempt fails with sufficient remaining time, i.e. the message is re-transmitted over the current hop for the $(k_s+1)$\textsuperscript{th} time.

On the other hand, for the right branch we have
\begin{equation}
%	\begin{split}
		S(R(\mathfrak{T}_s))
		=\begin{cases}
			\emptyset&t_s<T\subscript{min}^{(i_s)},\\
			\emptyset&i_s=0,\\
			\left\{(t_s-n_sT\subscript{b}-T\subscript{l}, i_s-1, 0, 0)\right\}&\text{o.w.}
		\end{cases}
%	\end{split}
\end{equation}
Again, the first case implies to insufficient remaining time and the second case describes a successful completion of \ac{DMH-HARQ}. The third case stands for a successful \ac{DF} over an intermediate hop, when the \ac{DMH-HARQ} progresses to the next hop, i.e. the $(i_s-1)$\textsuperscript{th} last hop.

\subsection{Optimization Problem}\label{subsec:op}
To measure the effectiveness of a \ac{DMH-HARQ} schedule, we define a utility $\xi(\mathfrak{T}_s)$ as the chance of successfully forwarding the message over the next $i_s$ hops within a time limit of $t_s$, then from the recursive structure {\eqref{eq:left_branch_binary_tree}--\eqref{eq:right_branch_binary_tree}} of $\mathfrak{T}_s$, we have
\begin{equation}\label{eq:recursive_utility}
	\xi(\mathfrak{T}_s)=\begin{cases}
		0&\mathfrak{T}_s=\emptyset,\\
		1&i_s=0,\\
		\varepsilon_s\xi(L(\mathfrak{T}_s))+(1-\varepsilon_s)\xi(R(\mathfrak{T}_s))&\text{o.w.},
	\end{cases}
\end{equation}
where $\varepsilon_s$ is the \ac{PER} of the $(k_s+1)$\textsuperscript{th} HARQ attempt on the $i_s$\textsuperscript{th} last hop with codeword blocklength $n_s$. Since $\varepsilon_s\in[0,1]$, we know that $\xi(\mathfrak{T}_s)$ is bounded between $[0,1]$ for any $\mathfrak{T}_s$.

Thus, given the time frame length $T\subscript{max}$ and total number of hops $I$ specified, the reliability of DMH-HARQ can be optimized regarding the blocklength allocation policy $n$:
\begin{maxi!}
	{n:\mathcal{S}\to\mathbb{N}^+}{\xi(\mathfrak{T}_{s_0})\label{prob:main}}{\label{prob:main_wo_con}}{}
	\addConstraint{n_{\mathrm{min},i_s}\leqslant n_s\leqslant \frac{t_s-T\subscript{min}^{(i_s-1)}}{Mf\subscript{s}}\quad\forall s\in\mathcal{S}\label{con:blocklength_limits}}
%	\addConstraint{n_s\in\mathbb{N}^+\quad\forall s\in\mathcal{S},\label{con:integer_blocklength}}
\end{maxi!}
where $s_0=(T\subscript{max},I,0,0)${, and the action $n_s=n(s)$ is the blocklength to be allocated for the next (re)transmission at state $s$}. The minimal blocklength $n_{\mathrm{min},i_s}$ is set here concerning the rapid decay of link reliability regarding reducing blocklength in the FBL regime. Commonly, people set an artificial constraint that the \ac{PER}  $\varepsilon_s$ for any $s\in\mathcal{S}$ shall not exceed a pre-defined threshold $\varepsilon\subscript{max}$, so that
\begin{equation}
	n_{\mathrm{min},i_s}=\min\{n\in\mathbb{N}^+~\vert~\varepsilon_s\leqslant\varepsilon\subscript{max}\}\label{eq:min_blocklength_general}
\end{equation}

{It shall be noted that \eqref{prob:main_wo_con} is \emph{not} a convex optimization problem since it is an integer programming problem. The existence (but not uniqueness) of optimal solution is guaranteed by the facts that \begin{enumerate*}[label=\emph{\roman*)}]
	\item it is a finite combinatorial optimization problem, and
	\item the objective function is limited within the range $[0,1]$.
\end{enumerate*}
In the worst-efficient approach, the optimal solution can be found by exhaustive search with an exponential time complexity. For a better efficiency, as we will show in Sec.~\ref{sec:design}, it can be solved within pseudo-polynomial time by means of integer \ac{DP}. Nevertheless, before starting with the integer \ac{DP} algorithm, some analyses to a relaxed convex version of Problem \eqref{prob:main_wo_con} can provide us useful insights to the structure of the optimal solution, which are usually crucial to the design of the integer \ac{DP} algorithm by means of reducing the state and action spaces to search. Such analyses will be provided in the next section.}

\section{Analyses: Type I HARQ without Combining}\label{sec:analyses}
Before diving deep into the optimal \ac{HARQ} scheduling problem, we shall study the error model of each \ac{HARQ} attempt. For a delay-critical scenario with stringent \ac{E2E} latency constraint, such as \ac{URLLC}, we consider the use of short code, where the blocklength of every codeword is limited. Therefore, the \ac{PER} of every \ac{HARQ} attempt shall be analyzed w.r.t. the finite blocklength information theory, and the performance highly depends on the selected coding and \ac{HARQ} schemes. 

{The choice of combining technique can have crucial impact in the performance of \ac{HARQ} protocols. It is well known that among the different types of \ac{HARQ}, Type I \ac{HARQ} without combining provides the worst error performance with the simplest implementation, while the Type II \ac{HARQ}-\ac{IR} offers the best error performance with the highest implementation complexity. On the one hand, applying Type II \ac{HARQ}-\ac{IR} to our \ac{DMH-HARQ} protocol will certainly grant it an optimal performance. Unfortunately, however, the generic error model will therewith become analytically intractable, due to the dual inconsistencies -- regarding both the redundancy rate and the frame length -- over individual transmission attempts. While numerical simulations can be conducted to obtain the empirical model dedicated to any certain channel coding scheme, such models lack generality and therefore have limited theoretic contribution. On the other hand, the Type I \ac{HARQ} without combining, known as the worst scheme regarding error performance, has a rich value in analysis since it outlines a tight upper-bound for \ac{PER} of all \ac{HARQ} solutions. Therefore, in this study we focus on the case of Type I \ac{HARQ} without  combining, where the receiver does not exploit any information of the unsuccessfully decoded packets.}

\subsection{Error Model, Constraints, and Approximations}\label{subsec:error_model_constraints}
Regarding Type I \ac{HARQ} without  combining, according to \cite{polyanskiy2010channel}, the \ac{PER} of an arbitrary \ac{HARQ} attempt with {payload length $d$ and} blocklength $m$ on the $i\superscript{th}$ hop is
\begin{equation}\label{eq:err_model_fbl}
	\epsilon_i(m)=Q\left(\sqrt{\frac{m}{V_i}}\left(\mathcal{C}_i-\frac{d}{m}\right)\ln2\right),
\end{equation}
where for \ac{AWGN} channels 
%\begin{equation}
	$V_i=1-\frac{1}{(1+\gamma_i)^2}${,}
%\end{equation}
%with $\gamma_i$ denoting the \ac{SNR} on the $i\superscript{th}$ hop:
%\begin{equation}
%	\gamma_i=\frac{Pg_i}{N}=\frac{P\overline{g}}{NL_{\text{f},i}}.
%\end{equation}
with $\gamma_i=\frac{Pg_i}{N}=\frac{P\overline{g}}{NL_{\text{f},i}}$ denoting the \ac{SNR} on the $i\superscript{th}$ hop. Here, $g_i$ is the channel gain of the $i\superscript{th}$ hop, which is determined by the mean channel gain $\overline{g}$ and the random fading loss $L_{\text{f},i}$. Thus, for all $s=(t_s, i_s, k_s, \tau_s)\in\mathcal{S}$, with the blocklength allocation policy $n_s$, we have
\begin{equation}\label{eq:err_model_harq_1_wo_sc}
%	\begin{split}
		\varepsilon_s=\epsilon_{I+1-i_s}(n_s)
		=Q\left(\sqrt{\frac{n_s}{V_{I+1-i_s}}}\left(\mathcal{C}_{I+1-i_s}-\frac{d}{n_s}\right)\ln2\right),
%	\end{split}
\end{equation}
which has no dependency on $t_s$, $k_s$ or $\tau_s$ except for the potential dependency of $n_s$ on them. 

%
%Thus, for arbitrary $\mathbf{m}_{i-1}$, we can obtain the probability that node $i$ successfully forwards the message with exactly $j$ HARQ attempts, where each $l\superscript{th}$ attempt takes a blocklength of $n_i\left(\mathbf{m}_i^{(l)}\right)$:
%\begin{equation}
%	\eta_{i,j}=\left\{1-\varepsilon_{i}\left[n_i\left(\mathbf{m}_i^{(j)}\right)\right]\right\}\prod\limits_{l=1}^{j-1}\varepsilon_{i}\left[n_i\left(\mathbf{m}_i^{(l)}\right)\right],\label{eq:trans_success_chance}
%\end{equation}
%where $j\in\{1,2,\dots k_i(\mathbf{m}_{i-1})\}$.

Following the common routine in FBL study, we can extend the space of blocklength from integer to real numbers, so as to allow non-integer values of $n_s$ for the convenience of analysis (approximate integer solutions can be then obtained by rounding the real-valued optima). Thus, the original
%constraint \eqref{con:integer_blocklength} can be removed.\footnote{As the relaxed one $n_s\in\mathbb{R}^+, \forall s\in\mathcal{S}$ is implicitly embedded in \eqref{con:blocklength_limits}.} 
problem \eqref{prob:main} can be replaced by
\begin{maxi}
	{n:\mathcal{S}\to\mathbb{R}^+}{\xi(\mathfrak{T}_{s_0})}{\label{prob:relaxed}}{}
\end{maxi}

Recalling \eqref{eq:min_blocklength_general},  the lower-bound $n_{\mathrm{min},i_s}$ in constraint \eqref{con:blocklength_limits} can be obtained from:
\begin{equation}
	Q\left(\sqrt{\frac{n_{\mathrm{min},i_s}}{V_{I+1-i_s}}}\left(\mathcal{C}_{I+1-i_s}-\frac{d}{n_{\mathrm{min},i_s}}\right)\ln2\right)=\varepsilon\subscript{max},
\end{equation}
which depends on $s$ only regarding $i_s$. Meanwhile, the upper-bound $n_{\mathrm{max},s}\triangleq\frac{t_s-T\subscript{min}^{(i_s-1)}}{Mf\subscript{s}}$ is determined by both $t_s$ and all $n_{\mathrm{min},j}$ that $1\leqslant j\leqslant i_s-1$, i.e., by $t_s$ and $i_s$.

\subsection{Multi-Hop Link Reliability with Perfect Global CSI}
Based on the error model above, we start our analysis to the optimization problem \eqref{prob:main_wo_con}  with a simplified case where perfect \ac{CSI} of all hops $i\in\mathcal{I}$ is available at every node.  For the convenience of notation, in the analyses below we let
\begin{equation}
	\xi_{\mathrm{max}}(s)\triangleq \max\limits_{n:\mathcal{S}\to\mathbb{R}^+}\xi(\mathfrak{T}_s),\qquad \forall s\in\mathcal{S},\label{eq:max_multi_hop_reliability}
\end{equation}
and the corresponding optimal blocklength allocation policy
\begin{equation}
	n\subscript{opt}(s)\triangleq \arg\max\limits_{n:\mathcal{S}\to\mathbb{R}^+}\xi(\mathfrak{T}_s),\qquad \forall s\in\mathcal{S}.\label{eq:opt_blocklength_alloc_policy}
\end{equation}

First, recall the remarks we made earlier in Section~\ref{subsec:error_model_constraints}, that given a certain $s$, both the \ac{PER} $\varepsilon_s$ and the constraints to $n_s$ are independent from $k_s$ or $\tau_s$, but determined by $t_s$ and $i_s$. Thus, we can assert that
\begin{lemma}\label{lemma:history_independence}
	With Type I \ac{HARQ} without combining, $n\subscript{opt}(s)$ is independent from $k_s$ or $\tau_s$, but determined by $t_s$ and $i_s$, i.e. $n\subscript{opt}(s)=n\superscript{opt}_{i_s}(t_s)$.
\end{lemma}
The proof of Lemma~\ref{lemma:history_independence} is trivial and therefore omitted.

For convenience we let  $s\subscript{l}(\mathfrak{T}_s)$ denote the root state of  $L(\mathfrak{T}_s)$, and similarly $s\subscript{r}(\mathfrak{T}_s)$ the root state of $R(\mathfrak{T}_s)$. Thus, from the recursive structure Eq.~\eqref{eq:recursive_utility} of the utility $\xi(\mathfrak{T}_s)$, for all $s$ that $\mathfrak{T}_s\neq\emptyset$ and $i_s\neq 0$, we can derive
\begin{equation}\label{eq:bellman}
	\begin{split}
		&\xi\subscript{max}(s)\\
		=&\max\limits_{n_s}\left\{\varepsilon_s\xi\subscript{max}(s\subscript{l}(\mathfrak{T}_s))+(1-\varepsilon_s)\xi\subscript{max}(s\subscript{r}(\mathfrak{T}_s))\right\}\\
		=&\max\limits_{n_s}\left\{\varepsilon_s\max\limits_{n_{S(L(\mathfrak{T}_s))}}\xi(L(\mathfrak{T}_s))+(1-\varepsilon_s)\max\limits_{n_{S(R(\mathfrak{T}_s))}}\xi(R(\mathfrak{T}_s))\right\}\\
		=&\dots
	\end{split}
\end{equation}
which forms a Bellman equation. Under the error model \eqref{eq:err_model_harq_1_wo_sc}, we have the following conclusions:

\begin{lemma}\label{lemma:one-shot}
%	As long as $\varepsilon\subscript{max}\leqslant 0.5$, 
	With Type I \ac{HARQ} without  combining, for all $s\in\mathcal{S}$ that $i_s=1$ (i.e. on the last hop), the optimal blocklength allocation is always a one-shot transmission $n\subscript{opt}(s)=\frac{t_s-T\subscript{l}}{T\subscript{b}}$ that maximizes the utility to $\xi\subscript{max}(s)=1-\epsilon_I\left(\frac{t_s-T\subscript{l}}{T\subscript{b}}\right)$.
\end{lemma}

\begin{theorem}\label{th:continuous_monotone_utility}
	As long as $t_s\geqslant T\subscript{min}^{(i_s-1)}$ and $i_s\geqslant 1$, $\xi\subscript{max}(s)$ is a continuous and strictly monotone increasing function of $t_s$.
\end{theorem}

% \begin{lemma}\label{lemma:concave_difference_left_right_branches}
% 	For all $s\in\mathcal{S}$, given a certain $n_s$, the function $\xi_\Delta(s)\triangleq\xi\subscript{max}(s\subscript{l}(\mathfrak{T}_s))-\xi\subscript{max}(s\subscript{r}(\mathfrak{T}_s))$ is non-positive, monotone increasing, and concave in wide sense.
% \end{lemma}

% {\begin{theorem}\label{th:unique_optimum}
% 	Forcing $n_s=0$ when $t_s<T\subscript{min}^{(i_s-1)}$ or $i_s=0$, with Type I \ac{HARQ} without combining, there is always one unique solution $n\subscript{opt}(s)$ to the Bellman equation \eqref{eq:bellman} for all $s\in\mathcal{S}$.
% \end{theorem}}

The proofs are provided in Appendices~\ref{app:proof_one-shot} and \ref{app:proof_continuous_monotone}, respectively.
% The proofs are provided in Appendices A and B, respectively.

\SetKwFunction{FOptSchdl}{OptSchedule}
\SetKwFunction{FBestUtlt}{Best!}
\SetKwFunction{FBestUtltSucc}{BestSucc!}
\SetKwFunction{FSchedule}{ScheduleTreeGen}
\SetKwProg{Pn}{function}{:}{}%
\begin{figure*}[!ht]
	%	\begin{alg}{Integer \ac{DP} for \ac{DMH-HARQ} schedule optimization}{}{}
	\removelatexerror
	\begin{algorithm}[H]
		\caption{Integer \ac{DP} for \ac{DMH-HARQ} schedule optimization}
		\label{alg:dmh-harq}
		\scriptsize
		\DontPrintSemicolon
		%			\SetKwFunction{FOptSchdl}{OptSchedule}
		%			\SetKwFunction{FBestUtlt}{Best!}
		%			\SetKwFunction{FBestUtltSucc}{BestSucc!}
		%			\SetKwFunction{FSchedule}{ScheduleTreeGen}
		%			\SetKwProg{Pn}{function}{:}{}
		
		\Pn{\FOptSchdl{$s, g_{i_s}, f_G, \varepsilon\subscript{max}, T\subscript{b}, T\subscript{l}$}}
		{
			$\mathcal{D}\subscript{local}\gets\mathrm{Dict}([~]), \mathcal{D}\subscript{succ}\gets\mathrm{Dict}([~])$\tcp*{Empty dictionaries.}
			\FBestUtlt{$t_s, i_s, g_{i_s}, f_G, \varepsilon\subscript{max}, \mathcal{D}\subscript{local}, \mathcal{D}\subscript{succ}$}\tcp*{Solve the optimal blocklength allocation policy.}
			$\mathfrak{T}_s\gets$\FSchedule{$t_s, i_s, 0, 0, \mathcal{D}\subscript{local}, T\subscript{b}, T\subscript{l}$}\tcp*{Generate the binary schedule tree.}
		}
		\Return$\mathfrak{T}_s$\;
		\;
		
		\Pn{\FBestUtlt{$t_s, i_s, g_{i_s}, f_G, \varepsilon\subscript{max}, \mathcal{D}\subscript{local}, \mathcal{D}\subscript{succ}$}}
		{
			%				\tcp{Two global dictionaries $\mathcal{D}\subscript{local}$ and $\mathcal{D}\subscript{succ}$ are used here as \ac{LUT}s for the best achievable utility and associated blocklength allocation: $\mathcal{D}\subscript{local}$ for the current hop with perfect \ac{CSI}, $\mathcal{D}\subscript{succ}$ to estimate successor hops with only statistical \ac{CSI}.}
			$n_{\mathrm{min},i_s}\gets\min\{n:\varepsilon_s(n)\leqslant \varepsilon\subscript{max}\}$\tcp*{W.r.t. Eq.~\eqref{eq:err_model_harq_1_wo_sc}.}
			$\overline{n}_{\mathrm{min}}\gets\min\{n:\overline{\varepsilon}(n)\leqslant \varepsilon\subscript{max}\}$\tcp*{W.r.t. Eq.~\eqref{eq:err_rate_imperfect_csi}.}
			\uIf{$\mathcal{D}\subscript{local}.\mathrm{find}\left([t_s,i_s]\right)\neq\emptyset$}{
				$[\xi\subscript{opt},n\subscript{opt}]\gets \mathcal{D}\subscript{local}.\mathrm{find}\left([t_s,i_s]\right)$\tcp*{Already solved, directly read from the \ac{LUT}.}
			}
			\uElseIf{$t_s<\left[(i_s-1)\overline{n}\subscript{min}+n_{\mathrm{min},i_s}\right]T\subscript{b}+i_sT\subscript{l}$}{
				$[\xi\subscript{opt},n\subscript{opt}]\gets[0,0]$\tcp*{Insufficient radio resource left, \ac{DMH-HARQ} terminated.}
				$\mathcal{D}\subscript{local}.\mathrm{append}\left([t_s,i_s]\rightarrow[\xi\subscript{opt},n\subscript{opt}]\right)$\tcp*{Append the new result to \ac{LUT}.}
			}
			\uElseIf{$i_s==1$}{
				$\left[\xi\subscript{opt}, n\subscript{opt}\right]\gets \left[\varepsilon_s(n\subscript{opt}), \left\lfloor\frac{t_s-T\subscript{l}}{T\subscript{b}}\right\rfloor\right]$\tcp*{Last hop, one-shot as optimum.}
				$\mathcal{D}\subscript{local}.\mathrm{append}\left([t_s,i_s]\rightarrow[\xi\subscript{opt},n\subscript{opt}]\right)$
			}
			\Else{
				\parbox[t]{.6\linewidth}{
					$n\subscript{opt}\gets \arg\max\limits_n\{\varepsilon_s(n)\FBestUtlt(t_s-nT\subscript{b}-T\subscript{l},i_s,g_{i_s},f_G,\varepsilon\subscript{max},\mathcal{D}\subscript{local},\mathcal{D}\subscript{succ})$\\
					$~~~~~~~~~~~+[1-\varepsilon_s(n)]\FBestUtltSucc(t_s-nT\subscript{b}-T\subscript{l},i_s-1,f_G,\varepsilon\subscript{max},\mathcal{D}\subscript{succ})\}$
				}\;
				\parbox[t]{.6\linewidth}{
					$\xi\subscript{opt}\gets\varepsilon_s(n\subscript{opt})\FBestUtlt(t_s-n\subscript{opt}T\subscript{b}-T\subscript{l},i_s,g_{i_s},f_G,\varepsilon\subscript{max},\mathcal{D}\subscript{local},\mathcal{D}\subscript{succ})+$\\
					$~~~~~~~~~~~[1-\varepsilon_s(n\subscript{opt})]\FBestUtltSucc(t_s-n\subscript{opt}T\subscript{b}-T\subscript{l},i_s-1,f_G,\varepsilon\subscript{max},\mathcal{D}\subscript{succ})$
				}\;
				$\mathcal{D}\subscript{local}.\mathrm{append}\left([t_s,i_s]\rightarrow[\xi\subscript{opt},n\subscript{opt}]\right)$
			}
		}
		\Return $\xi\subscript{opt}$\;
		\;
		
		\Pn{\FBestUtltSucc{$t_s, i_s, f_G, \varepsilon\subscript{max}, \mathcal{D}$}}
		{
			$\overline{n}_{\mathrm{min}}\gets\min\{n:\overline{\varepsilon}(n)\leqslant \varepsilon\subscript{max}\}$\;
			\uIf{$\mathcal{D}.\mathrm{find}\left([t_s,i_s]\right)\neq\emptyset$}{
				$[\xi\subscript{opt},n\subscript{opt}]\gets \mathcal{D}.\mathrm{find}\left([t_s,i_s]\right)$
			}
			\uElseIf{$t<i(\overline{n}\subscript{min}T\subscript{b}+T\subscript{l})$}{
				$[\xi\subscript{opt},n\subscript{opt}]\gets[0,0]$\;
				$\mathcal{D}.\mathrm{append}\left([t_s,i_s]\rightarrow[\xi\subscript{opt},n\subscript{opt}]\right)$
			}
			\uElseIf{$i_s==1$}{
				$\left[\xi\subscript{opt}, n\subscript{opt}\right]\gets \left[\varepsilon_s(n\subscript{opt}), \left\lfloor\frac{t_s-T\subscript{l}}{T\subscript{b}}\right\rfloor\right]$\;
				$\mathcal{D}.\mathrm{append}\left([t_s,i_s]\rightarrow[\xi\subscript{opt},n\subscript{opt}]\right)$
			}
			\Else{
				\parbox[t]{.6\linewidth}{
					$n\subscript{opt}\gets \arg\max\limits_n\{\overline{\varepsilon}(n)\FBestUtltSucc(t_s-nT\subscript{b}-T\subscript{l},i_s,f_G,\varepsilon\subscript{max},\mathcal{D})+$\\
					$~~~~~~~~~~~[1-\overline{\varepsilon}(n)]\FBestUtltSucc(t_s-nT\subscript{b}-T\subscript{l},i_s-1,f_G,\varepsilon\subscript{max},\mathcal{D})\}$
				}\;
				\parbox[t]{.6\linewidth}{
					$\xi\subscript{opt}\gets\overline{\varepsilon}(n)\FBestUtltSucc(t_s-nT\subscript{b}-T\subscript{l},i_s,f_G,\varepsilon\subscript{max},\mathcal{D})+$\\
					$~~~~~~~~~~~[1-\overline{\varepsilon}(n)]\FBestUtltSucc(t_s-nT\subscript{b}-T\subscript{l},i_s-1,f_G,\varepsilon\subscript{max},\mathcal{D})$
				}\;
				$\mathcal{D}.\mathrm{append}\left([t_s,i_s]\rightarrow[\xi\subscript{opt},n\subscript{opt}]\right)$
			}
		}
		\Return $\xi\subscript{opt}$\;
		\;
		
		\Pn{\FSchedule{$t_s, i_s, k_s, \tau_s, \mathcal{D}\subscript{local}, T\subscript{b}, T\subscript{l}$}}
		{
			\uIf{$\mathcal{D}\subscript{local}.\mathrm{find}([t_s,i_s])==\emptyset$}{
				$\mathfrak{T}_s\gets\emptyset$
			}
			\Else{
				$(\xi,n)\gets\mathcal{D}\subscript{local}.\mathrm{find}([t_s,i_s])$\;
				\parbox[t]{.6\linewidth}{
					$\mathfrak{T}_s\gets\left(\FSchedule(t_s-nT\subscript{b}-T\subscript{l}, i_s, k_s+1, \tau_s+nT\subscript{b}, \mathcal{D}\subscript{local}, T\subscript{b}, T\subscript{l}), \{(t_s, i_s, k_s, \tau_s)\},\right.$\\ $~~~~~~~~\left.\FSchedule(t_s-nT\subscript{b}-T\subscript{l}, i_s-1, 0, 0, \mathcal{D}\subscript{local}, T\subscript{b}, T\subscript{l})\right)$
				}
			}
		}
		\Return $\mathfrak{T}_s$
		\;		
	\end{algorithm}
	%	\end{alg}
% \end{strip}
\end{figure*}
\subsection{Multi-Hop Link Reliability with Imperfect CSI}
As we have discussed in Section~\ref{sec:setup}, in the multi-hop relaying scenario, it shall be considered that each node $i$ has only a perfect \ac{CSI} of the $i\superscript{th}$ hop. For all the successor hops $\{i+1,i+2,\dots I\}$, it possesses only the statistical \ac{CSI}, i.e. the \ac{PDF} $f_{G_i}$ of their channel gains $g_i$, which can be reasonably assumed consistent and identical for every hop $i\in\mathcal{I}$, i.e. $f_{G_i}\equiv f_G$ for all $i\in\mathcal{I}$. Thus, in an arbitrary state $s\in\mathcal{S}$ where the node $I+1-i_s$ is forwarding the message, the node cannot accurately estimate $\varepsilon_{s'}$ for any state $s'$ that $i_{s'}<i_s$ upon the blocklength allocation $n_{s'}$ anymore, but only the expectation upon the random channel gain:
\begin{equation}\label{eq:err_rate_imperfect_csi}
	\overline{\varepsilon}_{s'}=\int\limits_{0}^{+\infty}\varepsilon_{s'}f_G(g_{i_s'})\diff g_{i_{s'}}.
\end{equation}
Note that for any specific $\gamma_{i_{s'}}\geqslant 0$, $\varepsilon_{s'}$ is convex and monotone decreasing w.r.t. $n_{s'}$. Thus, as a linear combination of various samples of $\varepsilon_{s'}$ over $\gamma_{i_{s'}}\in[0,+\infty)$, $\overline{\varepsilon}_{s'}$  is also convex and monotone decreasing w.r.t. $n_{s'}$, so that Theorem~\ref{th:continuous_monotone_utility} remains applicable even when only statistical \ac{CSI} of for all successor hops is available at each node.

\section{Algorithm Design}\label{sec:design}
\subsection{Optimal DMH-HARQ: Integer Dynamic Programming}
From Lemma~\ref{lemma:history_independence} we know that when applying Type I \ac{HARQ} without combining, the \ac{DMH-HARQ} scheduling is a typical \ac{MDP}, where the expected utility $\xi_s$ from any certain status $s$ is independent from its historical states or actions, but only relying on the remaining time $t_s$ and the blocklength allocation policy $n_s$. Such kind of problems can be generally solved by recursive approaches that gradually evaluate and improve the decision policy by traversing the entire feasible region of states, typical examples are policy iteration and value iteration. However, such approaches are challenged by two issues when dealing with Eq.~\eqref{eq:bellman}. First, \eqref{eq:err_model_harq_1_wo_sc} provides no closed-form analytical solution $n_s$ for an arbitrary $\varepsilon_s$. Second, with $n_s\in\mathbb{R}^+$, both the feasible region $\mathcal{S}$ and the set of possible actions becomes infinite, making it impossible to traverse the solution space within limited time. 

Nevertheless, it shall be remarked that the real-value relaxation $n_s\in\mathbb{R}^+$ was taken, like in classical FBL works, only for the convenience of analysis. Indeed, representing the blocklength per transmission in the unit of channel use, the final solution $n\subscript{opt}(s)$ can only take integer values from $\mathbb{N}^+$. Meanwhile, to apply FBL approaches while guaranteeing to meet the stringent latency requirement, as referred earlier, $n_s$ is strictly constrained by \eqref{con:blocklength_limits}. Thus, both the action set and the feasible region of status are usually of reasonable sizes, enabling us to apply classical integer \ac{DP} techniques to directly solve the global optimum {(the feasibility and optimality of this approach are discussed in \cite{han2022clarq})}. A typical implementation is to recursively solve $n\subscript{opt}(s)$ from the last hop where $i_s=1$ on, hop-by-hop onto the first one where $i_s=I$. The recursive algorithm is essentially accompanied with \ac{LUT} to store the optimal achievable utility over the solution space, as described by Algorithm~\ref{alg:dmh-harq}:
\begin{itemize}
	\item Two global dictionaries, $\mathcal{D}\subscript{local}$ and $\mathcal{D}\subscript{succ}$, are defined in the main optimizing function \FOptSchdl as \ac{LUT}s for the solutions of local hop (based on perfect \ac{CSI}) and successor hops (based on statistical \ac{CSI}), respectively. Each \ac{LUT} maps the state space $\mathcal{S}$ onto $\mathbb{R}^+$ to record the achievable utility $\xi\subscript{max}(s)$.
	\item The functions \FBestUtlt and \FBestUtltSucc are implemented to recursively solve the optimal blocklength allocation policy $n_s$ and the associated best utility $\xi\subscript{max}(s)$ regarding perfect and imperfect CSI, respectively. Especially, \FBestUtltSucc is a deviated version of \FBestUtlt that considers with the only statistical \ac{CSI} for each hop, while \FBestUtlt exploits the perfect \ac{CSI} of the local hop. \FBestUtltSucc	is called by \FBestUtlt to estimate the achievable utility over successor hops. \emph{The global dictionaries dictionary $\mathcal{D}\subscript{local}$ and $\mathcal{D}\subscript{succ}$ are therewith updated, respectively}. 
	\item The function \FSchedule is implemented to recursively generate the binary schedule tree $\mathfrak{T}$ from the dictionary $\mathcal{D}\subscript{local}$ that stores the blocklength allocation policy.
\end{itemize}

\subsection{Hop-by-Hop CSI Update and Reschedule}\label{subsec:hop-by-hop}
Note that by running Algorithm~\ref{alg:dmh-harq} at node $i\in\mathcal{I}$, the optimal schedule $\mathfrak{T}_s$ is solved based on the perfect CSI of the $i\superscript{th}$ hop and statistical CSI of all hops thereafter, which shall not be adopted by any other node. Indeed, upon a successful decoding of the message received from its predecessor node, each node $i\in\mathcal{I}$ shall individually solve its own optimal schedule $\mathfrak{T}_s$ regarding the instantaneous system state $s$ and its real-time channel measurement $g_i$ of the $i\superscript{th}$ hop. The complete procedure can be briefly summarized as in Algorithm~\ref{alg:hop-by-hop}. \revise{Note that the total number of hops $I$ is not endogenous to the algorithm, but only passed to the first node to initialize the state $s$. This allows our approach to be flexibly applied in the practical scenarios, where the number of hops is dynamically determined by the specific service function chain and the network topology.}

%\begin{alg}{$I$-hop \ac{DMH-HARQ} with \ac{CSI} update}{}{}
\removelatexerror
\begin{algorithm}
	\caption{$I$-hop \ac{DMH-HARQ} with \ac{CSI} update}
	\label{alg:hop-by-hop}
	\footnotesize
	\DontPrintSemicolon
	
	Initialization: $[g_1, g_2, \dots g_I], f_G, \varepsilon\subscript{max}, T\subscript{max}, T\subscript{b}, T\subscript{l}$\;
	$s\gets(T\subscript{max},I,0,0)$\;
	\For{$i=I, I-1, \dots, 1$}{
		$\mathfrak{T}_s\gets$\FOptSchdl{$s,g_i,f_G,\varepsilon\subscript{max}, T\subscript{b}, T\subscript{l}$}\;
		$s'\gets s$\;
		\While{true}{
			(Re)-transmit regarding $\mathfrak{T}_s$\;
			Update $s$ regarding the transmission result\;
			\uIf{$n_{s}=0$}{
				\Return\tcp*{Termination state}
			}
			\ElseIf{$i_{s}<i_{s'}$}{
				\textbf{break}\tcp*{Current hop succeeded}
			}
		}
	}
	\Return{}
\end{algorithm}
%\end{alg}

\subsection{Computational Complexity Analysis}\label{subsec:complexity}
As described above, Algorithm~\ref{alg:dmh-harq} implements a recursive \ac{DP} approach to compute the integer dynamic program. The recursive algorithm considers the \ac{DMH-HARQ} as an $I$-stage \ac{DP} problem, each stage representing a hop. Meanwhile, each hop-level sub-problem that starting with state $s$ makes also an $\mathcal{O}(n_{\mathrm{max},s})$-stage \ac{DP} problem itself, each stage representing an available channel use in the schedule. 
{With any limited $T_{\max}$, maximum of level of stage is finite and upper-bounded, i.e., $I_{\max}\revise{\leqslant}\frac{T_{\max}}{Mf_s n_{\min,i_s}}$, where $n_{\min,i_s}$ is the minimum
blocklength defined in~\eqref{eq:min_blocklength_general}. Therefore, the convergence of the algorithm is guaranteed. }

According to the computational complexity analysis of the \ac{CLARQ} algorithm~\cite{han2022clarq}, which has the same structure as each hop-level sub-problem, we know that the time complexity of each hop-level sub-problem is $\mathcal{O}(n_{\mathrm{max},s})$. Since for each $s\in\mathcal{S}$ hop, $n_{\text{max},s}$ is upper-bounded by $n\subscript{MAX}\triangleq\frac{t_s-T\subscript{min}^{(I-1)}}{Mf\subscript{s}}$, the time complexity of solving the optimal DMH-HARQ at the $i\superscript{th}$ last node is ${\mathcal{O}\left(n\subscript{MAX}^i\right)}$. Taking the hop-by-hop execution of Algorithm~\ref{alg:hop-by-hop} into account, the overall time complexity of the DMH-HARQ approach over the $I$-hop relay chain is $\mathcal{O}\left[\sum\limits_{i=1}^In\subscript{MAX}^i\right]=\mathcal{O}\left(n\subscript{MAX}^I\right)$, \revise{which is significantly higher than to optimize the schedule of conventional static \ac{HARQ} approaches, i.e. $\mathcal{O}\left[\log \left(n\subscript{MAX}\right)\right]$.}

In realistic scenarios of deployment, such a time complexity can critically challenge the online computation of optimal \ac{DMH-HARQ} schedule, especially due to the time-varying channel gains $g_i$, which leads to high delay and significant power consumption. To address this issue, it is a practical solution to implement a \ac{LUT} in each node, which that contains a set of offline solved dictionaries regarding various channel conditions{, i.e., the integer \ac{DP} algorithm is pre-executed offline, only the \acp{LUT} must be deployed at different relaying nodes}.  Since the operation of searching for a specific entry in a dictionary-based LUT has only a time complexity of $\mathcal{O}(1)$, the devices can be rapidly adapted to the appropriate specification w.r.t. the real-time channel measurement, leading to an overall time complexity as low as $\mathcal{O}(I)$ for the entire $I$-hop \ac{DMH-HARQ} process.

\section{Numerical Evaluation}\label{sec:simulation}
To evaluate our proposed approaches, we conducted numerical simulations to benchmark \ac{DMH-HARQ} against baseline solutions with respect to different specifications of the system and the environment.

\begin{table*}[!htpb]
	\centering
	\caption{Setup of the simulation campaign}
	%	\begin{tabular}{llm{3.5cm}}
		\begin{tabular}{llm{8cm}}
			\toprule[2px]
			\textbf{Parameter}			&	\textbf{Value}				&	\textbf{Remark}\\
			\midrule[1.5px]
			$f\subscript{s}$			&	$250$kSPS					&	Symbol rate, corresponding to a \SI{4}{\micro\second} symbol duration\\% OFDM symbol length in 5G numerology 4\\
			{$Q\subscript{m}$}							&	{$1$}							&	Modulation order, {1} for BPSK\\
			$P$							&	\SI{1}{\milli\watt}			&	Transmission power\\
			$N$							&	\SI{1}{\milli\watt}			&	Noise power\\
			$d$							&	\SI{16}{\bit}				&	Payload of each message\\
			$\varepsilon\subscript{max}$&	$0.5$						&	Maximal allowed PER per transmission\\
			$T\subscript{max}$			&	\SI{1}{\milli\second}		&	Default time frame length \\
			$\mathbf{T}\subscript{max}$	&	$\{0.4,0.5,\dots,1\}$~\si{\milli\second}	&	Test range of time frame length\\
			$I$							&	$4$							&	Default number of hops\\
			$\mathbf{I}$				&	$\{2,3,\dots,10\}$				&	Test range of hops number\\
			$L_{\text{f},i}$				&	$\sim\mathrm{Rice}(0.5,1)$	&	Small-scale fading with a consistent \ac{LoS} path\\
			$T\subscript{l}$			&	$\SI{12}{\micro\second}$	&	Default decoding/feedback delay\\
			$\mathbf{T}\subscript{l}$	&	$\{10,20,\dots,100\}$~\si{\micro\second}	&	Test range of decoding/feedback delay\\
			$\overline{g}$				&	\SI{0}{\dB}					&	Mean channel gain\\
			$\overline{\mathbf{g}}$		&	$\{-10,-9,\dots,0\}$~\si{\dB}	&	Test range of mean channel gain\\
			$n\subscript{MC}$			&	${1~000~000}$					&	Runs per Monte-Carlo test\\
			\bottomrule[2px]
		\end{tabular}
		\label{tab:sim_setup}
	\end{table*}

\subsection{Baseline Methods}\label{subsec:baselines}
Three baseline methods are implemented and evaluated in our simulations:
\begin{enumerate}
	\item {\bf Naive static \ac{HARQ}{-}\ac{IR}}: where the (re)transmission times on each hop and coding rate of every individual (re)transmission are pre-scheduled. For each hop, {a full \ac{IR} (Type II \ac{HARQ})} is applied to combine the codes sent in different transmissions. Since the performance is highly dependent on the (re)transmission times and the coding rate, here we consider the approximate bound provided by \emph{Makki} et al. in \cite{makki2014finite}, which is only achievable with ideal Gaussian codebooks allowing non-integer blocklength. Under a naive schedule, the same length of time is assigned to every hop. {Recall that the performance of Type I \ac{HARQ} is guaranteed outperformed by that of Type II, and that our \ac{DMH-HARQ} approach is evaluated with Type I \ac{HARQ}. Therefore, the benchmark is unfairly favoring the baseline over our approach.}
	
	\item {\bf Optimal static \ac{HARQ}-\ac{IR}}: which follows the same principle of pre-scheduled retransmission as the naive static \ac{HARQ}-\ac{IR}. However, the sub-frame lengths allocated to different hops are jointly optimized so as to minimize the \ac{E2E} packet loss rate. {Again}, Type II \ac{HARQ}-\ac{IR} is applied and the bound in \cite{makki2014finite} is considered, {granting this baseline an unfair advantage in competition against our approach}.
	
	\item {\bf Listening-based cooperative \ac{ARQ}}: which was proposed by \emph{Goel} et al. in \cite{goel2021listen}. {More specifically, here we consider its fully-cumulative scheme which outperforms other variants.} Similar to the naive static \ac{HARQ} approach, every hop is assigned with the same length of sub-frame and pre-scheduled with the same number of (re)transmissions and coding rate. However, if the \ac{DF} on some hop $i$ succeeds before using up all the pre-scheduled (re)transmission slots assigned to it, the \ac{DF} on hop $i+1$ will immediately be triggered, inheriting all the remaining time resource from its predecessor hop. In our simulation, each hop is pre-scheduled with $2$ (re)transmission slots, applying Type-{I} \ac{HARQ} without combining.
\end{enumerate}

\begin{figure}[!htbp]
	\centering
	\includegraphics[width=.9\linewidth]{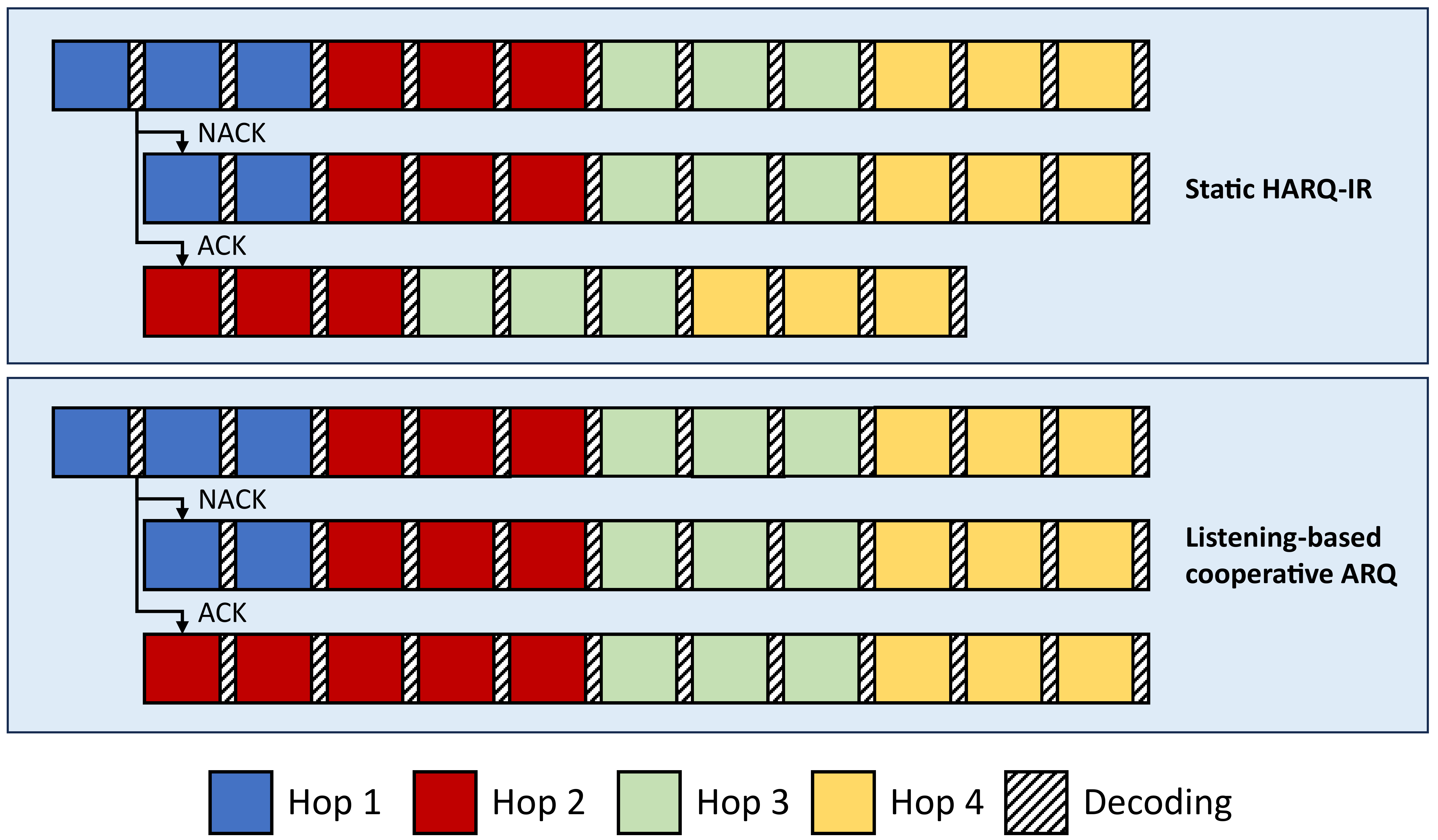}
	\caption{{Illustration of baseline solutions for 4-hop \ac{DF} chain.}}
	\label{fig:baselines}
\end{figure}

The principles of both static \ac{HARQ}-\ac{IR} and listening-based cooperative \ac{ARQ} are illustrated in Fig.~\ref{fig:baselines} for comparison with Fig.~\ref{fig:dmh-harq}. Note that the naive and optimal schemes of static \ac{HARQ}-\ac{IR} only distinguish from each other by {whether} the sub-frames are of of same length over all hops.

\subsection{Simulation Setup}\label{subsec:sim_setup}
The setup of our simulation campaign is detailed in Tab.~\ref{tab:sim_setup}. {It is worth remarking that we consider here the E2E delay $T\subscript{max}$ no longer than \SI{1}{\milli\second}. Considering the low mobility of transceivers of wireless backhaul channels, this setup ensures that the delay over each single hop is lower than the coherence time of \ac{LoS} channels in the centimeter wave range~\cite{RBC2009spatial,LKT+2021positioning}, which is commonly used for such channels.} Among all the parameters, there are four {that are of particular interest to us} to test the sensitivity of our approach, namely the total time frame length $T\subscript{max}$, the number of hops $I$, the decoding/feedback delay $T\subscript{l}$, and the mean channel gain $\overline{g}$. Each of these key parameters is assigned with a default value and a test range. When benchmarking the approaches regarding one key parameter over its test range, all the rest key parameters are specified to their default values. For every individual specification, $n\subscript{MC}={1~000~000}$ runs of Monte-Carlo test are conducted, each run with an independently generated set of channel conditions for the $I$ hops, with which all approaches are tested.

\subsection{Results and Analyses}\label{subsec:results}
\begin{figure*}[!htpb]
	\centering
	\begin{subfigure}{.49\linewidth}
		\centering
		\includegraphics[width=\linewidth]{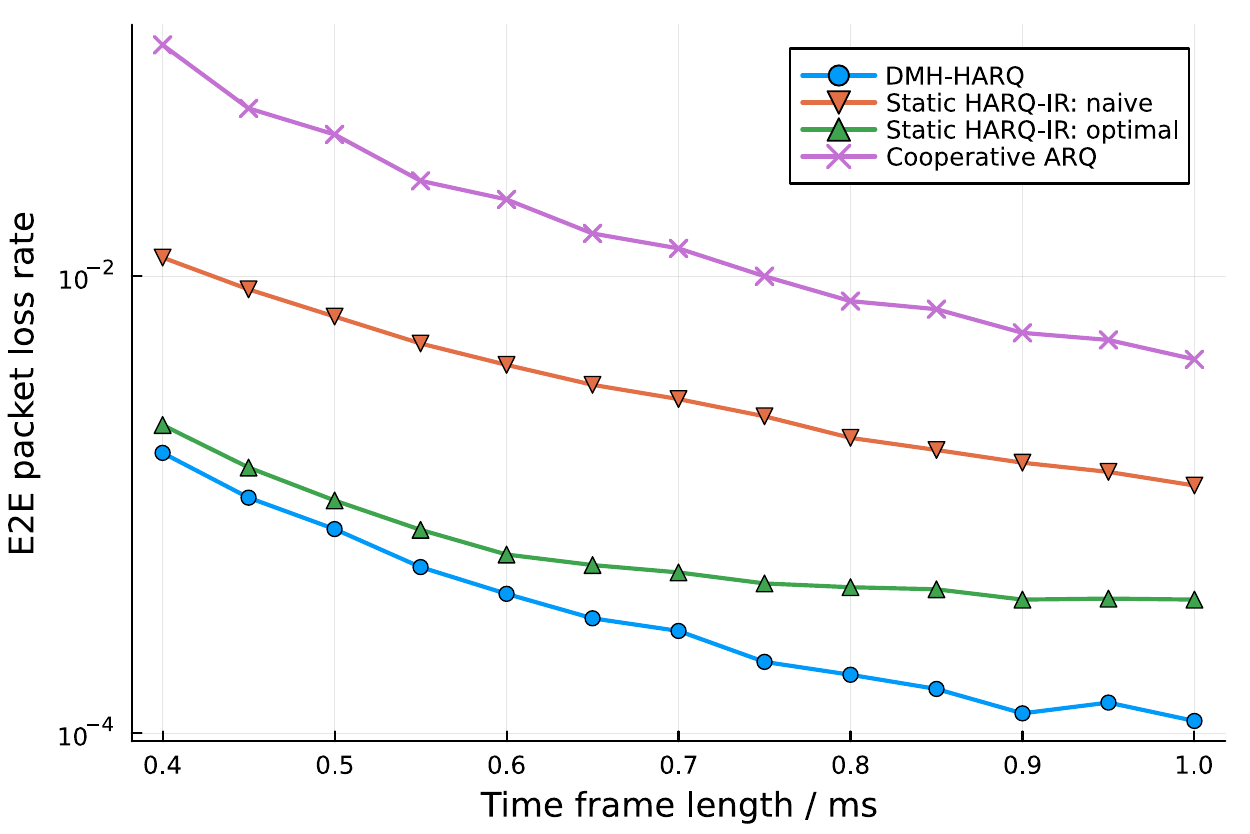}
		\subcaption{}
		\label{subfig:frame_length_test}
	\end{subfigure}
	\begin{subfigure}{.49\linewidth}
		\centering
		\includegraphics[width=\linewidth]{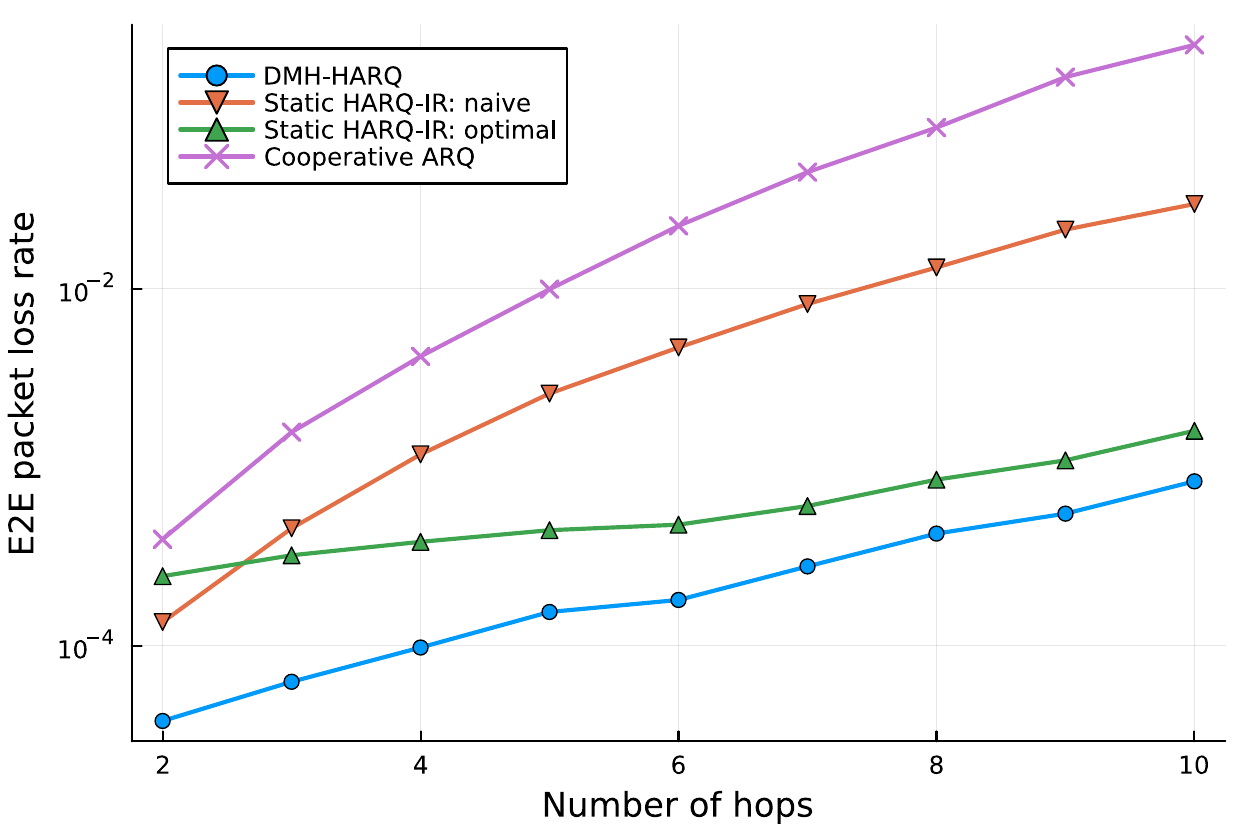}
		\subcaption{}
		\label{subfig:num_hops_test}
	\end{subfigure}
	\begin{subfigure}{.49\linewidth}
		\centering
		\includegraphics[width=\linewidth]{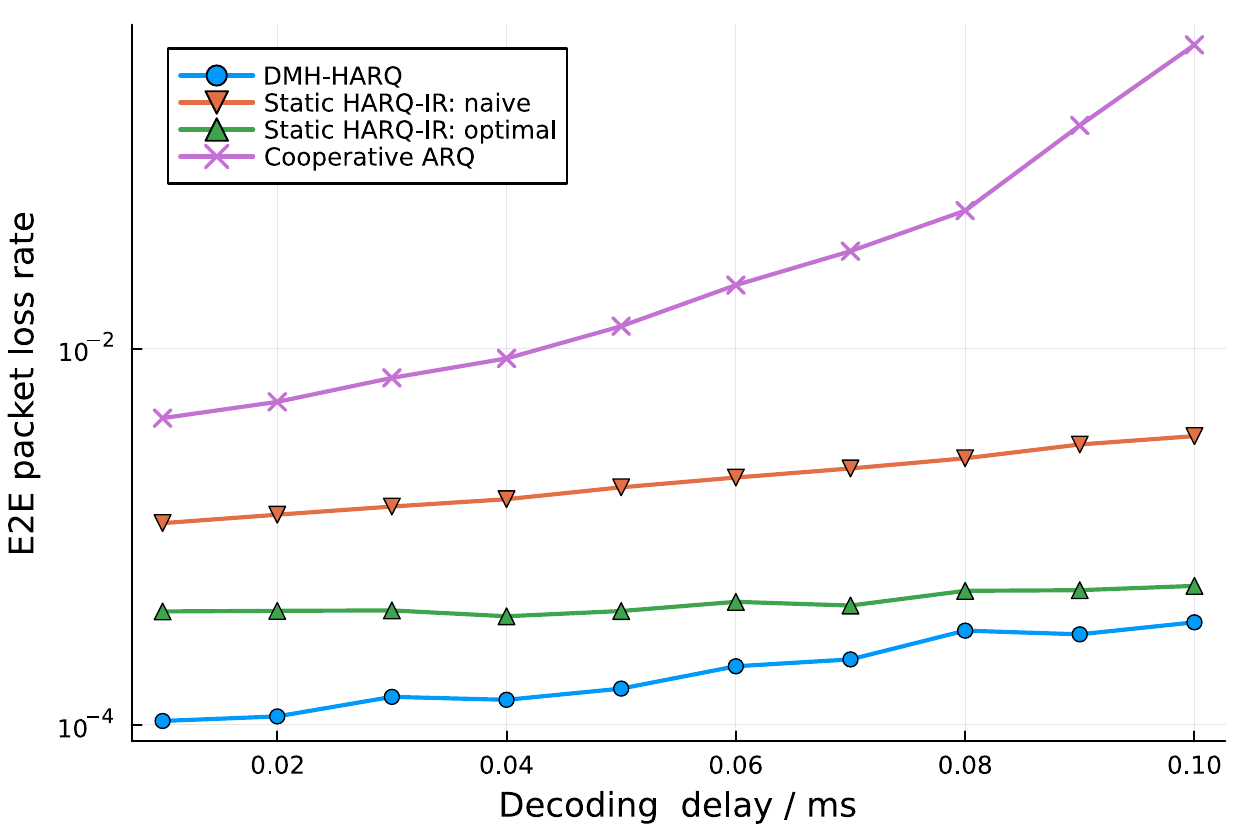}
		\subcaption{}
		\label{subfig:dec_delay_test}
	\end{subfigure}
	\begin{subfigure}{.49\linewidth}
		\centering
		\includegraphics[width=\linewidth]{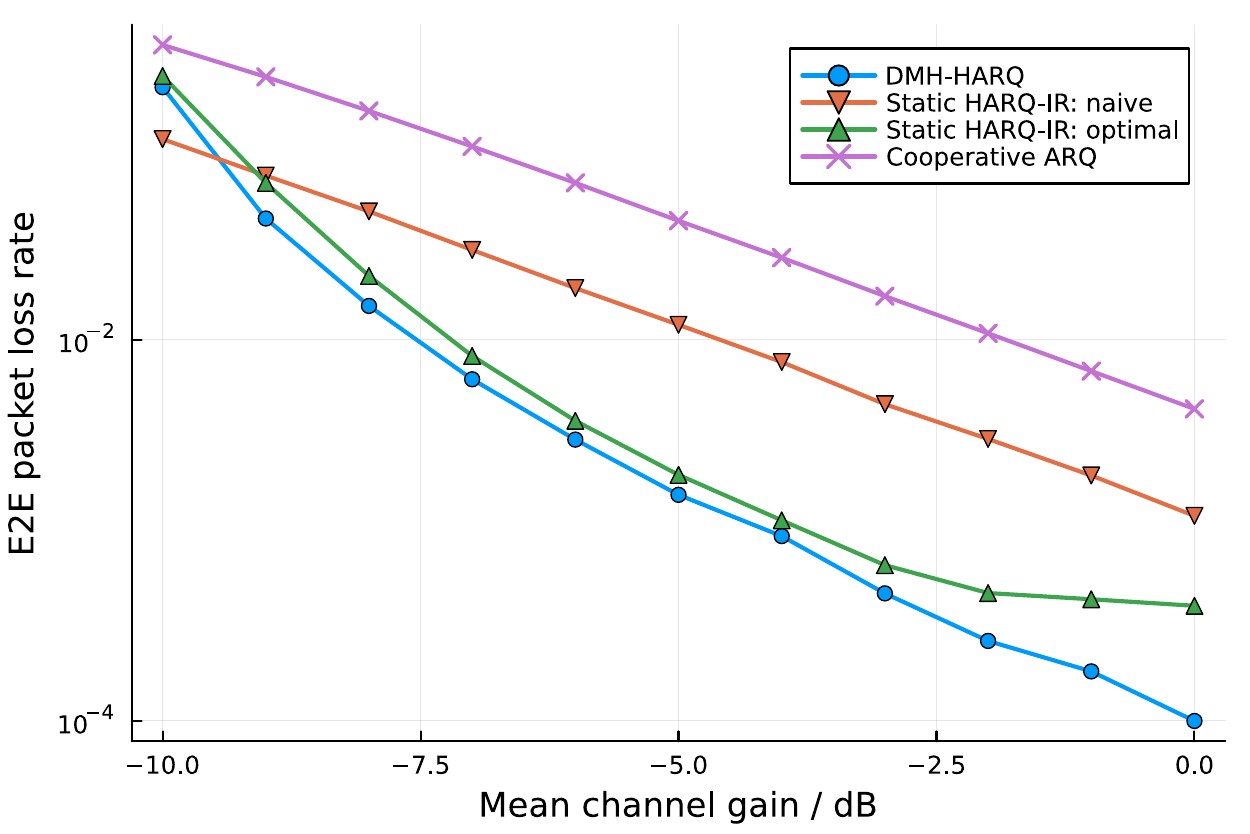}
		\subcaption{}
		\label{subfig:path_loss_test}
	\end{subfigure}
	\caption{{Sensitivity tests regarding (\subref{subfig:frame_length_test}) the time frame length, (\subref{subfig:num_hops_test}) the number of hops, (\subref{subfig:dec_delay_test}) the decoding/feedback delay, and (\subref{subfig:path_loss_test}) the pathloss, respectively.}}
	\label{fig:benchmarks}
\end{figure*}

The results of sensitivity tests regarding $T\subscript{max}$, $I$, $T\subscript{l}$ and $\overline{g}$ are illustrated in {Fig.~\ref{fig:benchmarks}}. Generally, the E2E packet loss rate monotone increase as \begin{enumerate*}[label=\emph{\roman*)}]
	\item the average blocklength that can be allocated to each hop  (exclusive the decoding/feedback delay) decreases; and
	\item the SNR {decreases}.
\end{enumerate*}

We observe from the results that the \ac{DMH-HARQ} approach consistently outperforms all baselines, {even when under an unfair disadvantage in soft combining. It is promised to hold even more performance gain over the static \ac{HARQ} solutions when applied with a more advanced combining technology. This excellent performance} is credited to the \ac{DP} algorithm that guarantees to achieve the theoretical maximum of the Bellman sum~\eqref{eq:bellman}. Especially, with more available blocklength per hop, since the overall possible numbers of \ac{HARQ} attempts over the $I$-hop chain increases, this performance gain brought by \ac{DP} is also increasing. To the contrary, when the channel is harsh or the radio resource is extremely limited, the constraint~\eqref{con:blocklength_limits} is likely forcing to allow only one transmission slot per hop, so that the optimal \ac{DMH-HARQ} schedule converges to the optimal one-shot-per-hop schedule, \revise{which} just gives the performance upper bound of optimal static \ac{HARQ}-\ac{IR} in \ac{FBL}~\cite{makki2014finite}. \revise{Moreover, the performance gain of \ac{DMH-HARQ} over the baselines remains significant when the number of hops increases.}

Remark that the listening-based cooperative \ac{ARQ} approach also allocates the radio resource to different hops in a technically dynamic manner. However, its mechanism design only allows the successor hops to inherit unused resource from predecessors, not the other way around. With a same fixed blocklength for every (re)transmission, it results in a longer effective sub-frame for the late hops that are closer to the final information sink than the for early hops that are closer to the source. However, as it can be derived from \cite{bellman1954some}, as a stochastic \ac{DP}, the optimal solution shall be just the opposite: to allocate more resource to the early hops than the late ones. In the end, the listening-based cooperative \ac{ARQ} approach cannot hit the optimum of the \ac{DP} problem~\eqref{prob:main_wo_con}, and has only a limited gain from the dynamic scheduling. This impact is even more critical in the \ac{FBL} regime where \ac{ARQ}/\ac{HARQ} generally work poor, leading to a significant gap to the performance in the \ac{IBL} regime. As we see from the results, the listenting-based cooperative \ac{ARQ} approach performs the worst among all baselines.

\section{Further Discussions}\label{sec:discussion}
% {
% 	\subsection{Computational Complexity}
% 	Concerning the recursive nature of Algorithm~\ref{alg:dmh-harq}, its computational load can be a crucial factor in assessing its practical applicability. Since \ac{DMH-HARQ} can be considered as an $I$-hop generalization of the $2$-hop \ac{CLARQ} protocol, we can follow the approach we used in \cite{han2022clarq} to analyze its time complexity, which returns us a result of $\mathcal{O}\left(n\subscript{MAX}^I\right)$. 
	
% 	It must be noted that, though this time complexity is significantly higher than to optimize the schedule of conventional static \ac{HARQ} approaches, i.e. $\mathcal{O}\left[\log \left(n\subscript{MAX}\right)\right]$, the optimizer is not necessarily executed in real time for every communication session. Indeed, the optimal schedules regarding different \ac{CSI} can be obtained offline for every node with Algorithm~\ref{alg:dmh-harq}, and cumulatively stored in a \ac{LUT}. In the runtime, the node only needs to look up the table for its optimal decision, which has a minimal time complexity of $\mathcal{O}\left(1\right)$.
% }

\begin{figure}[!htpb]
	\centering
	\revisebox{
	\includegraphics[width=.9\linewidth]{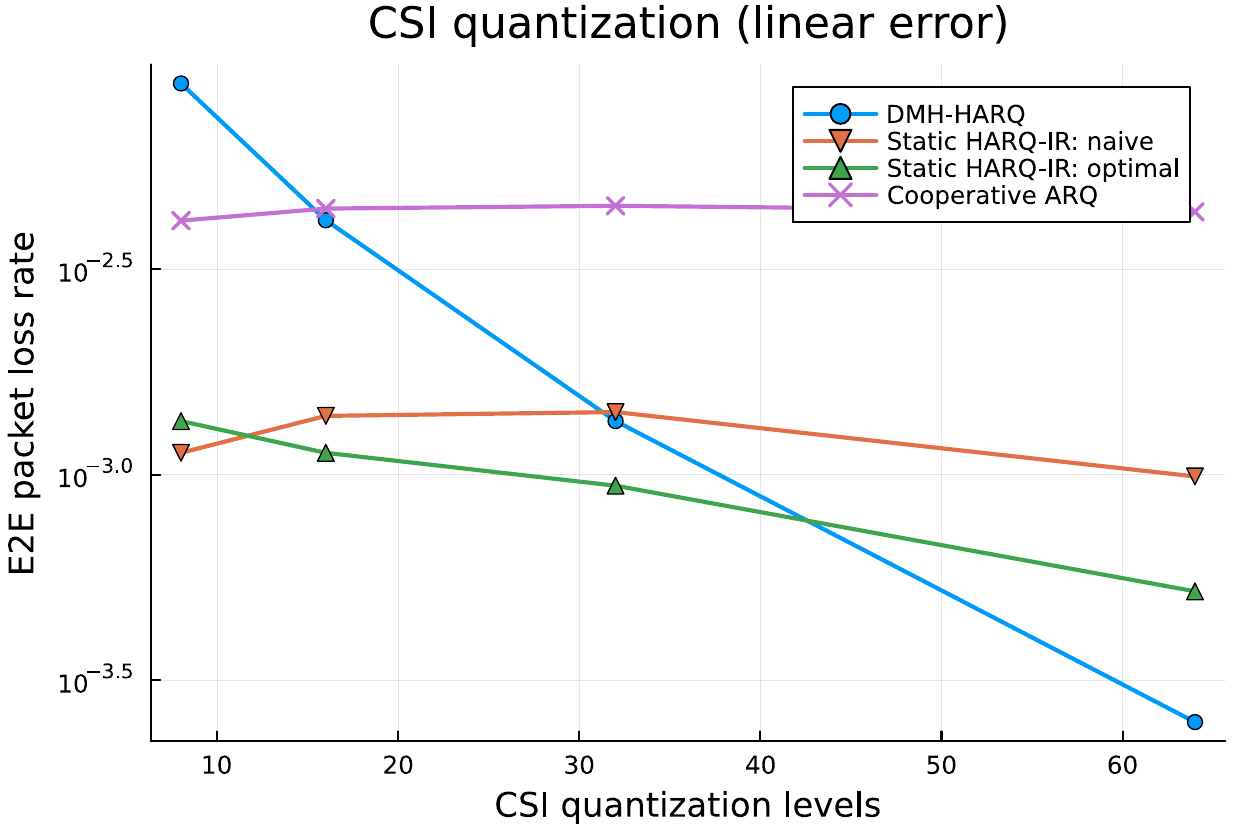}}
	\captionsetup{font={color=\highlightcolor}}
	\caption{\revise{Sensitivity test regarding the \ac{CSI} quantization levels.}}
	\label{fig:csi_quant_test}
\end{figure}

\revise{
\subsection{LUT Implementation: Complexity and Performance}\label{subsec:lut_cost}
As we have discussed in Sec.~\ref{subsec:complexity}, the requirement of real-time online solution suggests a \ac{LUT}-based implementation of the \ac{DMH-HARQ} method. As the price for the significant time complexity reduction from $\mathcal{O}(n\subscript{MAX}^I)$ to $\mathcal{O}\left[\log((n\subscript{MAX}))\right]$, extra space complexity must be taken into account. As we have already shown, for any given SNR $\gamma$, the allocation policy $n\subscript{opt}$ is a function of remaining time $t$ and the number of remaining hops $i$. Thus, the size of the \ac{LUT} implemented at each relaying node is (loosely) upper-bounded by $I\subscript{max}n\subscript{MAX}L$, where $n\subscript{MAX}$ is the maximum number of blocklength in a radio frame, and $L$ is the number of discrete SNR levels. Here we carry out a worst-case analysis for a rough estimate of the memory cost to assess the deployment feasibility. 

$I\subscript{max}$, as aforementioned, is constrained by the \ac{E2E} latency limit $T\subscript{max}$, which is typically in the range of tens to hundered milliseconds in our investigated use scenarios. Moreover, for every specific deployment, $I$ is determined by the overall distance of wireless backhaul and the radio coverage of each relay node, which dependents on the frequency band. For long-distance backhaul where rural environments are commonly considered, the traditional sub-\SI{40}{\giga\hertz} band is preferred and its typical link length (per hop) is around \SI{15}{\kilo\meter}. For midhaul connections in suburban areas, this length reduces to \SI{8}{\kilo\meter} per hop~\cite{HHG+2020rainfall}. According to \cite{gomes2019reducing}, relay chains in midhauls can reach a distance of \SI{40}{\kilo\meter} and in backhauls up to \SI{300}{\kilo\meter}, we can reasonably consider that $I\subscript{max}\leqslant 25$ even in the utmost case. $n\subscript{MAX}$, on the other hand, depends on the specific radio frame design for the wireless transport network, and is also constrained by the latency limit $T\subscript{max}$. At last, the \ac{SNR} quantization levels $L$%, we take the \ac{CQI} table size in \ac{5G} \ac{NR}, which is $16$, as a reasonable reference~\cite{3gppts38214}.
can be flexibly configured. However, a small $L$ implies higher \ac{CSI} quantization error, which will certainly degrade the performance. To assess this effect and select the proper $L$, we repeated the benchmark test with default specifications listed in Tab.~\ref{tab:sim_setup}, but with $L$-level quantized \ac{CSI} instead of accurate \ac{CSI} at each node. The results are illustrated in Fig.~\ref{fig:csi_quant_test}, which suggests to pick $L\geqslant 64$, so that significant performance loss can be avoided to sufficiently leverage the performance gain of \ac{DMH-HARQ}.

Given these parameter values, to provide a realistic estimate of the time and memory costs of the \ac{LUT} implementation, we benchmarked the computation of $L=64$ \ac{LUT}s regarding Algorithm~\ref{alg:dmh-harq} under different scenarios. The system specifications and the benchmark results are listed in Tab.~\ref{tab:lut_complexity}, which shows that even for an extremely challenging scenario with $I\subscript{max}=25$, $n\subscript{MAX}=5000$, and $d=128$, the offline \ac{LUT} computation can be accomplished within $5.5$ hours on a regular commercial workstation, and the total size of the generated \acp{LUT} is barely above \SI{250}{\kilo\byte}, which is negligible for modern communication devices.
% Assuming a $250$-symbol frame, the LUT size is loosly upper-bounded by $25\times 16\times 250=100~000$ entries with one $8$-bit integer per entry, which requires an insignificant memory cost of \SI{800}{\kilo\byte}. This burden, even if further increased regarding the quantization levels or the frame length by an order of magnitude, is still effortlessly affordable for a relay node in practical wireless transport networks.
}

\begin{table}[!htpb]
	\centering
	\captionsetup{font={color=\highlightcolor}}
	\caption{Cost of the offline \ac{LUT} generation ($L=64$)}
	\label{tab:lut_complexity}
	\revise{
		\begin{tabular}{m{3mm}|>{\centering\arraybackslash}m{4.2cm}>{\centering\arraybackslash}m{2.9cm}}
			\toprule[2px]
			&	\textbf{CPU}	&	\textbf{Memory}\\
			&	AMD~Ryzen~5~5600G @ \SI{3.9}{\giga\hertz} &	\SI{32}{\giga\byte}\\
			&	6 cores, 12 threads	&DDR4-\SI{2133}{\mega\hertz}\\
			&	\SI{64}{\bit}, \SI{7}{\nano\meter} process	&	Dual-channel\\
			&&\\
			&	\textbf{Operation System}	&	\textbf{\ac{LUT} Solver}\\
			\multirow{-7}{*}{\rotatebox{90}{\textbf{Testing Platform}}}&	Windows~11~Pro  &	Julia 1.11.6\\
			\midrule[1px]
		\end{tabular}
		\begin{tabular}{>{\centering\arraybackslash}m{4.6cm}>{\centering\arraybackslash}m{1.4cm}>{\centering\arraybackslash}m{1.4cm}}
			\textbf{Scenario}	&	\textbf{Time}	&	\textbf{Memory}\\
			$n\subscript{MAX}=100,~I\subscript{max}=2,~d=16$ &	\SI{464.8}{\milli\second}	&	\SI{4.876}{\kilo\byte}\\
			$n\subscript{MAX}=500,~I\subscript{max}=4,~d=32$ &	\SI{31.21}{\second}	&	\SI{27.62}{\kilo\byte}\\
			$n\subscript{MAX}=1000,~I\subscript{max}=10,~d=64$ &	\SI{4.021}{\minute}	&	\SI{48.20}{\kilo\byte}\\
			$n\subscript{MAX}=5000,~I\subscript{max}=25,~d=128$ &	\SI{5.467}{\hour}	&	\SI{250.88}{\kilo\byte}\\
			\bottomrule[2px]
		\end{tabular}
	}
\end{table}

\revise{\subsection{Imperfect CSI Measurement at Local Hop}\label{subsec:imperfect_csi}
While we have assumed that each node has perfect \ac{CSI} at its local hop and statistical \ac{CSI} at all hops thereafter, it shall be noted that in practical systems, even the \ac{CSI} measurement at the local hop can be inaccurate or outdated, and some performance degradation will occur in this case, in addition to the impact of the \ac{CSI} quantization error. Nevertheless, for wireless backhaul and midhaul scenarios where both the transmitter and receiver are mounted at high altitudes with stable positions, the channel variation is generally slow, and the \ac{CSI} can be accurately estimated with low overhead. Moreover, it is not only our proposed \ac{DMH-HARQ} scheme that will be affected by such imperfect \ac{CSI}. Conventional solutions, such like the static \ac{HARQ}-\ac{IR}, also rely on \ac{CSI} to perform optimal blocklength allocation. It is therefore of generic interest to minimize the \ac{CSI} estimation error and its impact on system performance, which is beyond the scope of this article.}

\revise{\subsection{Timer Overhead}\label{subsec:timer_overhead}
While the \ac{DMH-HARQ} approach requires a timer field embedded in the message header, it is also worth a discussion to assess its overhead. The timer field must contain sufficient bits to represent the remaining symbols in the current multi-hop forwarding session, so it is lower-bounded by $\lceil\log_2(n\subscript{MAX})\rceil$. We consider the same four scenarios as listed in Tab.~\ref{tab:lut_complexity}, and calculate the timer-to-payload bit length ratio for each, as shown in Tab.~\ref{tab:timer_overhead}. Generally, for small data packets over short relay chains, the overhead is significant but acceptable compared to the typical channel coding redundancy, considering the reliability gain it offers. Moreover, this overhead ratio further decreases in long-distance scenarios with larger packets, since the timer grows only logarithmically with $n\subscript{MAX}$, while decreasing propotionally to the payload length.}

\begin{table}[!htpb]
	\centering
	\captionsetup{font={color=\highlightcolor}}
	\caption{Timer-to-payload length ratio of selected scenarios}
	\label{tab:timer_overhead}
	\revise{
		\begin{tabular}{c c c}
			\toprule[2px]
			\textbf{Scenario}	&	\textbf{Min. Timer Bits}	& \textbf{Ratio}\\
			\midrule[1px]
			$n\subscript{MAX}=100,~d=16$ 	&	4	&	25.00\%\\
			$n\subscript{MAX}=500,~d=32$ 	&	9	&	28.13\%\\
			$n\subscript{MAX}=1000,~d=64$ 	&	10	&	15.63\%\\
			$n\subscript{MAX}=5000,~d=128$ 	&	13	&	10.16\%\\
			\bottomrule[2px]
		\end{tabular}
	}
\end{table}

\subsection{Incremental Redundancy Gain}
We have proven that even when applied without any combining technique, \ac{DMH-HARQ} is able to outperform the optimal static \ac{HARQ}-\ac{IR}. A natural idea, of course, is  to further enhance the performance of \ac{DMH-HARQ} with soft combining, especially with an ideal \ac{IR}, where the \ac{PER} formula~\eqref{eq:err_model_harq_1_wo_sc} must be replaced by
\begin{equation}\label{eq:err_model_harq_ir}
%	\begin{split}
		\varepsilon_s=Q\left(\sqrt{\frac{n_s+\frac{\tau_s}{T\subscript{b}}}{V_{I+1-i_s}}}\left(\mathcal{C}_{I+1-i_s}-\frac{d}{n_s+\frac{\tau_s}{T\subscript{b}}}\right)\ln2\right).
%	\end{split}
\end{equation}
Though \ac{IR} is guaranteed to reduce the achievable \ac{E2E} packet loss rate of \ac{DMH-HARQ}, it costs more computational effort. The inclusion of $\tau_s$ into $\varepsilon_s$ rejects Lemma~\ref{lemma:history_independence}. To keep it an \ac{MDP} so that the integer \ac{DP} based framework of Algorithm~\ref{alg:dmh-harq} still applies, $\tau_s$ must play its role as part of the system state when solving the decision policy, i.e.  $n_s$.  The dictionaries $\mathcal{D}\subscript{local}$ and $\mathcal{D}\subscript{succ}$ in Algorithm~\ref{alg:dmh-harq} shall map $[t_s,i_s,\tau_s]$ instead of $[t_s, i_s]$ to $[\xi\subscript{opt}, n\subscript{opt}]$. This will significantly increase the effective size of the state space, raising the time complexity of solving $n_s$ at each node from $\mathcal{O}(n\subscript{MAX}^I)$ to $\mathcal{O}(n\subscript{MAX}^{2I})$.

\subsection{Multi-Access Design}
%\todo[bh]{Revise/extend this section to: 1) explain why we exclude the 6G air interface from this; 2) discuss the TDMA-based multi-access design at CU/DU; and 2) explain that the algorithm can also be adopted for other multi-access systems if the air interface constraint is relaxed (e.g. using TDMA instead of OFDMA)}
In this paper we have been focusing on a single multi-hop data link. In practical deployment, a multi-access solution will be mandatory to allow efficient networking{,} especially in the fronthaul and midhaul domains, where the network exhibits a star topology that each \ac{O-DU} is associated with multiple \glspl{O-RU} and each \ac{O-CU} with multiple \glspl{O-DU}. While \ac{OFDMA} is widely used in modern radio systems and has been dominating most wireless standards, it may not be the optimal match for our \ac{DMH-HARQ} protocol, due to the challenge in interference management caused by the dynamic traffic pattern that are hard to predict. Similar issue has already been discussed for the original \ac{CLARQ} problem~\cite{han2022clarq}, revealing that \ac{TDMA} will be a better solution than \ac{OFDMA} in this context. This will limit the application of the dynamic \ac{HARQ} approaches such like \ac{CLARQ} or \ac{DMH-HARQ} for the air interface between users and base stations, which are generally standardized in modern cellular systems to use \ac{OFDMA} for better scalability and flexibility. It is also for this reason that we exclude the air interface from the scope of this study.

However, in wireless transport networks which is under our focus in this study, \ac{TDMA} can still be a good and feasible option, because: \begin{enumerate*}[label=\emph{\roman*)}]
	\item unlike the air interface of \ac{RAN}, the physical layer design of transport network is not universally standardized but upon the design of vendor, allowing to use \ac{TDMA};
	\item compared to the number of \glspl{UE} in \ac{RAN}, the number of transport network nodes are limited, leading to low demand for scalability; 
	\item the transport network nodes are generally installed at fixed position, leading to easy synchronization and low demand for flexibility.
\end{enumerate*}
%\todo{revise}
 
% Certainly, for some WLAN and PLAN radio access technologies that support \ac{TDMA}, such as Bluetooth or IEEE 802.16, 
 
%In practical scenarios such like \ac{IIoT} and \ac{V2X}, multiple such data links usually coexist in the same radio environment, or even share some of the relaying nodes sometimes, leading to cross-user interference that degrades the \ac{SINR}. While modern radio technologies are commonly integrated with interference management solutions to mitigate strong interference, most of the popular methods will fail in the context of \ac{DMH-HARQ}, since they rely on a prior knowledge of the scheduled transmission slots of \ac{UE}s and a prediction of the interfering traffic pattern. The mechanism of dynamically making decisions for (re)transmissions that are not aligned with fixed time frames, however, makes it impossible to make such predictions accurately. It becomes therefore a more appropriate design to use \ac{TDMA} rather than \ac{OFDMA} for \ac{DMH-HARQ}.

Moreover, the \ac{TDMA} design in \ac{DMH-HARQ} can be potentially combined with the \ac{TWT} mechanism, which is utilized by the IEEE 802.11ax standard, to further raise the time efficiency of \ac{DMH-HARQ}: after successfully forwarding a message, the relaying node can send a triggering token to the next transmitting node, so that it is immediately waken from sleep and start its the \ac{DMH-HARQ} process, instead of waiting till the last transmitting node completing its multi-hop relay. In such way, multiple data links can be parallelized in a pipeline fashion over the multiple hops, so that the \ac{E2E} delay is reduced.

\section{Conclusion and Outlooks}\label{sec:conclusion}
In this paper, we have looked into the problem of dynamic \ac{HARQ} in multi-hop wireless transport network with a special focus on reliability and openness. Considering \ac{FBL} regime, we have proposed a \ac{DMH-HARQ} scheme with an associated \ac{DP} algorithm that optimizes it. The proposed methods are proven via numerical simulations as effective and outperforming conventional baselines, even without any code combining. {Its performance superiority remains consistent regarding the link delay budget and decoding delay, and increases along with the hop number and the channel quality.}

As for the next step, we plan to apply \ac{IR} to \ac{DMH-HARQ}, and extend our analyses and optimization methods in this paper thereto. Deeper study to the multiple access design for \ac{DMH-HARQ} and its integration with \ac{TWT} mechanism is also an interesting topic for future study.

%\ifCLASSOPTIONcaptionsoff
%  \clearpage
%\fi

%\bibliographystyle{IEEEtran}
%\bibliography{references}


%\vfill

\begin{IEEEbiography}[{\includegraphics[width=1in,height=1.25in,clip,keepaspectratio]{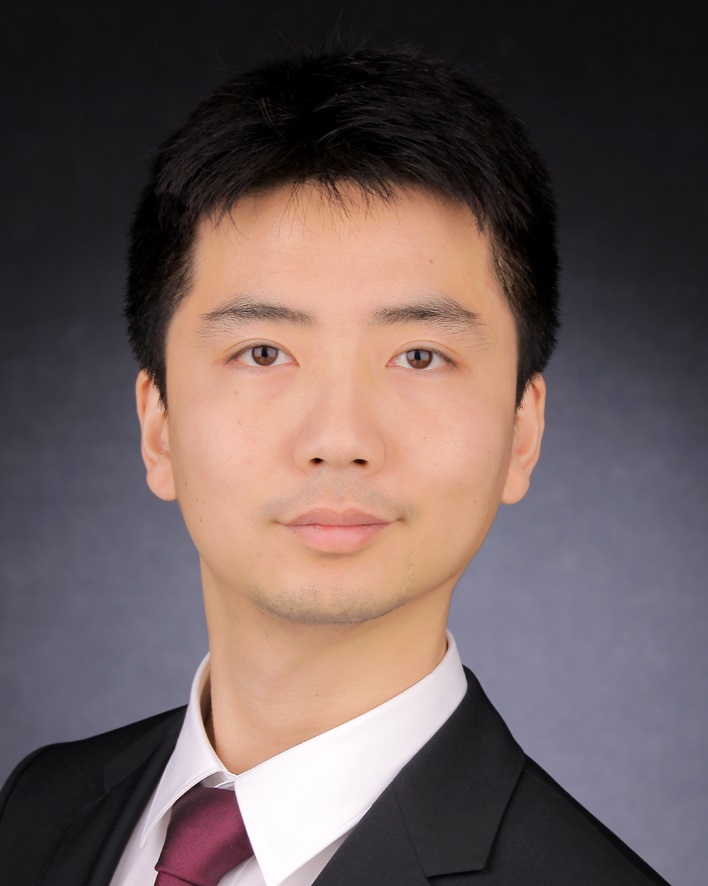}}]{Bin Han} (Senior Member, IEEE) received his B.E. degree in 2009 from Shanghai Jiao Tong University, M.Sc. in 2012 from the Technical University of Darmstadt, and a Ph.D. degree in 2016 from Karlsruhe Institute of Technology. He joined TU Kaiserslautern (which became later RPTU Kaiserslautern-Landau) in 2016, and was granted his Habilitation (Venia Legendi) in 2023. His research interests are in the broad areas of wireless communications, mobile networks, and signal processing. He is the author of two books, six book chapters, and over 90 research papers. He has participated in multiple EU FP7, Horizon 2020, Horizon Europe, and German BMBF/BMFTR research projects. Dr. Han is an Editor of \emph{IEEE Wireless Communications Letters} and an Editorial Board Member of \emph{Network}. He has served in organizing committee and/or TPC for \emph{IEEE GLOBECOM}, \emph{IEEE ICC}, \emph{EuCNC}, \emph{European Wireless}, and \emph{ITC}. He is actively involved in the IEEE Standards Association Working Groups P1955, P2303, P3106, and P3454.
\end{IEEEbiography}
%\vfill

\begin{IEEEbiography}[{\includegraphics[width=1in,height=1.25in,clip,keepaspectratio]{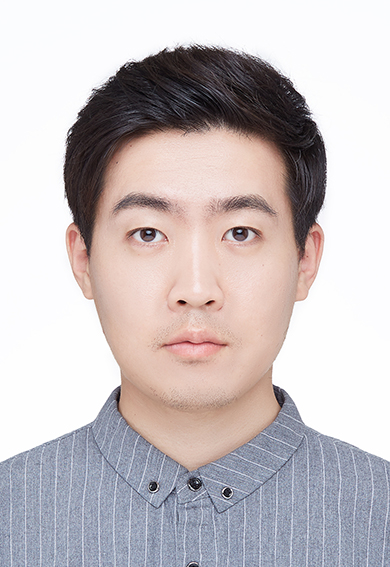}}]{Muxia Sun} (Member, IEEE) received in 2010 his B.Sc. degree from South China University of Technology (SCUT), M.Sc. in 2012 \& 2013 from Universit\'e de Nantes and SCUT, respectively, and the Ph.D. degree in 2019 from Universit\'e Paris-Saclay. Since 2020 he has been with Tsinghua University as Postdoctoral Researcher in the Department of Industrial Engineering. His current research interests include reliability assessment and optimization of industrial \& communication systems, robust optimization, and approximation algorithm design.
\end{IEEEbiography}
%\vfill

\begin{IEEEbiography}[{\includegraphics[width=1in,height=1.25in,clip,keepaspectratio]{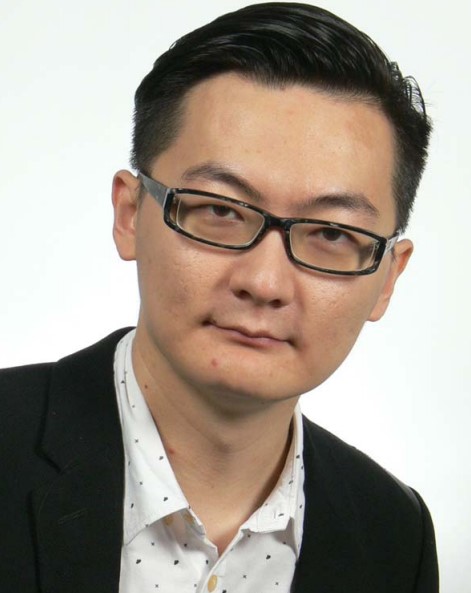}}]{Yao Zhu} (Member, IEEE) received the B.S. degree in electrical engineering from the University of Bremen, Bremen, Germany, in 2015. He received the M.Sc. degree and the Ph.D. degree (summa cum laude) in information technology and computer engineering from RWTH Aachen University, Aachen, Germany, in 2018 and 2022, respectively. He was a Post-Doctoral Research Fellow at RWTH Aachen University from 2022 to 2025. He is currently a Professor with the School of Electronic Information, Wuhan University, Wuhan China. His research interests include ultra-reliable and low-latency communications, mobile edge networks and physical layer security.
\end{IEEEbiography}
%\vfill

\begin{IEEEbiography}[{\includegraphics[width=1in,height=1.25in,clip,keepaspectratio]{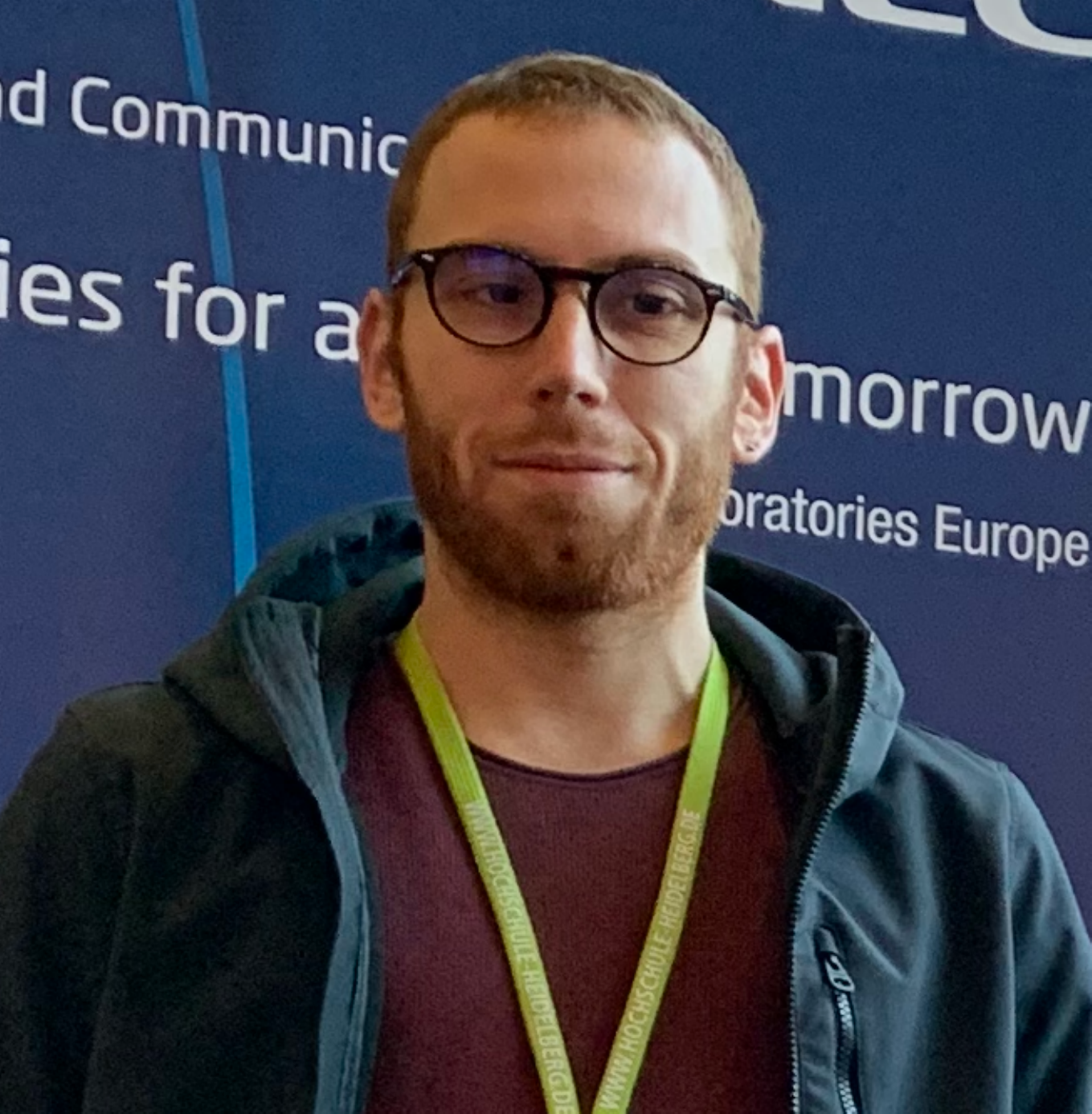}}]{Vincenzo Sciancalepore} (Senior Member, IEEE) received his M.Sc. degree in Telecommunications Engineering and Telematics Engineering in 2011 and 2012, respectively, whereas in 2015, he received a double Ph.D. degree. Currently, he is a senior 5G researcher at NEC Laboratories Europe GmbH in Heidelberg, focusing his activity on Smart Surfaces and Reconfigurable Intelligent Surfaces (RIS). He is currently involved in the IEEE Emerging Technologies Committee leading the initiatives on RIS. He was also the recipient of the national award for the best Ph.D. thesis in the area of communication technologies (Wireless and Networking) issued by GTTI in 2015. He is an Editor of \emph{IEEE Transactions on Wireless Communications} and Editor of \emph{IEEE Transactions on Communications}.
\end{IEEEbiography}
%\vfill

\begin{IEEEbiography}[{\includegraphics[width=1in,height=1.25in,clip,keepaspectratio]{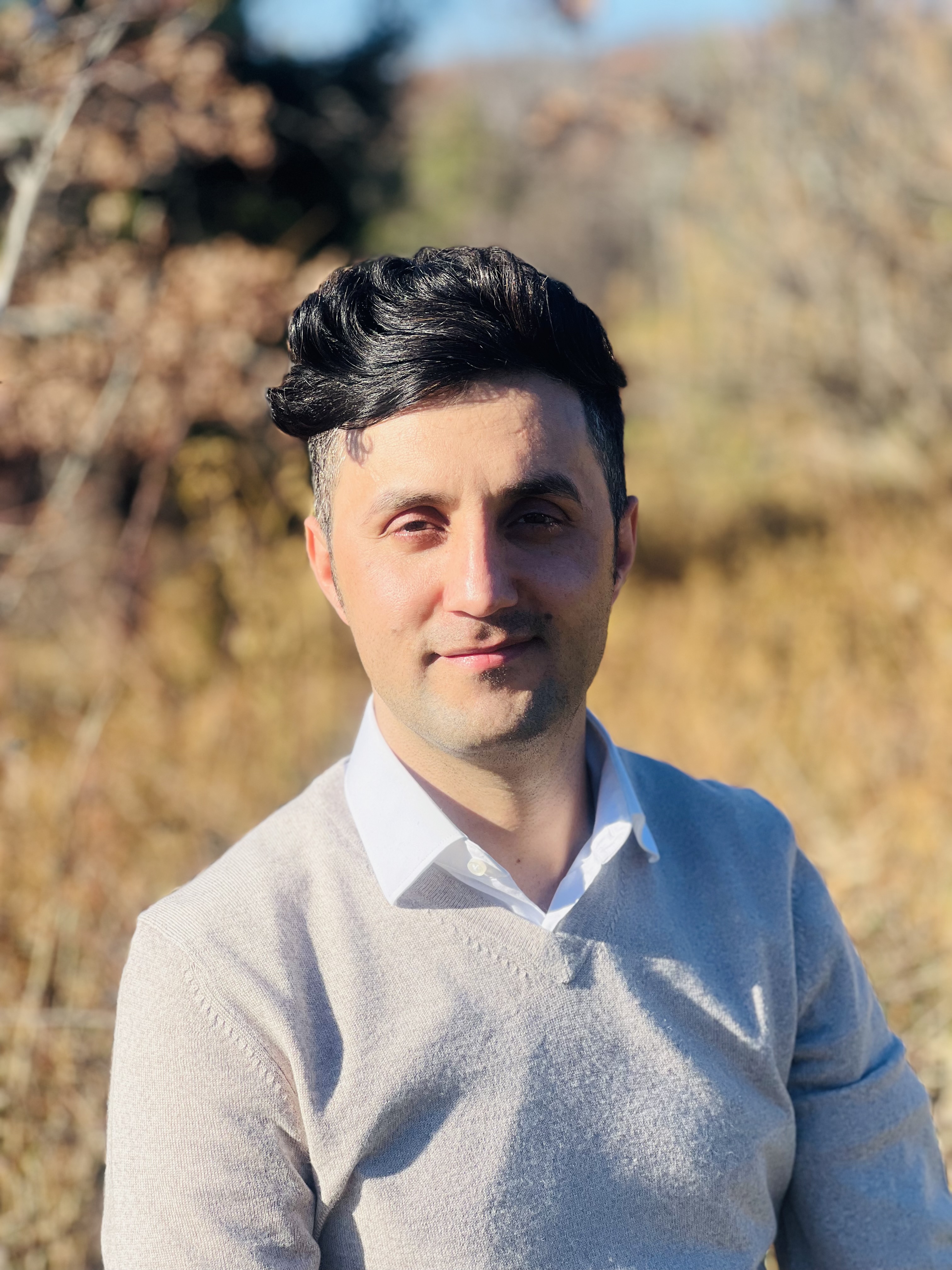}}]{Mohammad Asif Habibi} received his B.Sc. degree in Telecommunications Engineering from Kabul University, Afghanistan, in 2011. He obtained his M.Sc. degree in Systems Engineering and Informatics from the Czech University of Life Sciences, Czech Republic, in 2016. Since January 2017, he has been working as a research fellow and Ph.D. candidate at the Division of Wireless Communications and Radio Positioning, Rheinland-Pf\"alzische Technische Universit\"at Kaiserslautern-Landau (previously known as Technische Universit\"at Kaiserslautern), Germany.  From 2011 to 2014, he worked as a radio access network engineer for HUAWEI. His main research interests include network slicing, network function virtualization, resource allocation, machine learning, and radio access network architecture.
\end{IEEEbiography} 
%\vfill

\begin{IEEEbiography}[{\includegraphics[width=1in,height=1.25in,clip,keepaspectratio]{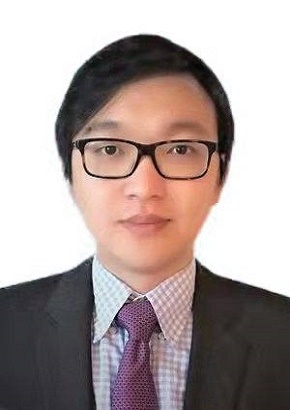}}]{Yulin Hu}(Senior Member, IEEE) received the Ph.D.E.E. degree (Hons.) from RWTH Aachen University.  In 2016, he was a Post-Doctoral Fellow in RWTH Aachen University, where he was a Senior Researcher with Prof. Anke Schmeink at ISEK institute. From May 2017 to July 2017, he was a Visiting Scholar with Prof. M. Cenk Gursoy with Syracuse University, USA. He is currently a Professor with the School of Electronic Information, Wuhan University. His research interests are in information theory and optimal design of wireless communication systems. His work received the Best Paper Awards at IEEE ISWCS 2017 and IEEE PIMRC 2017, and was listed as a Best Paper candidate at ACM MSWiM 2022. He served as a TPC member for many conferences including ICC, Globecom, WCNC, and served as the WS\&SS Chair for IEEE SmartData 2022, the Track Co-Chair for ICCCN 2023, and for Special Sessions in IEEE ISWCS 2018, 2021, and 2023. He is currently serving as Editor for journals including IEEE Transactions on Mobile Computing and IEEE Transactions on Vehicular Technology.
\end{IEEEbiography}
%\vfill

\begin{IEEEbiography}[{\includegraphics[width=1in,height=1.25in,clip,keepaspectratio]{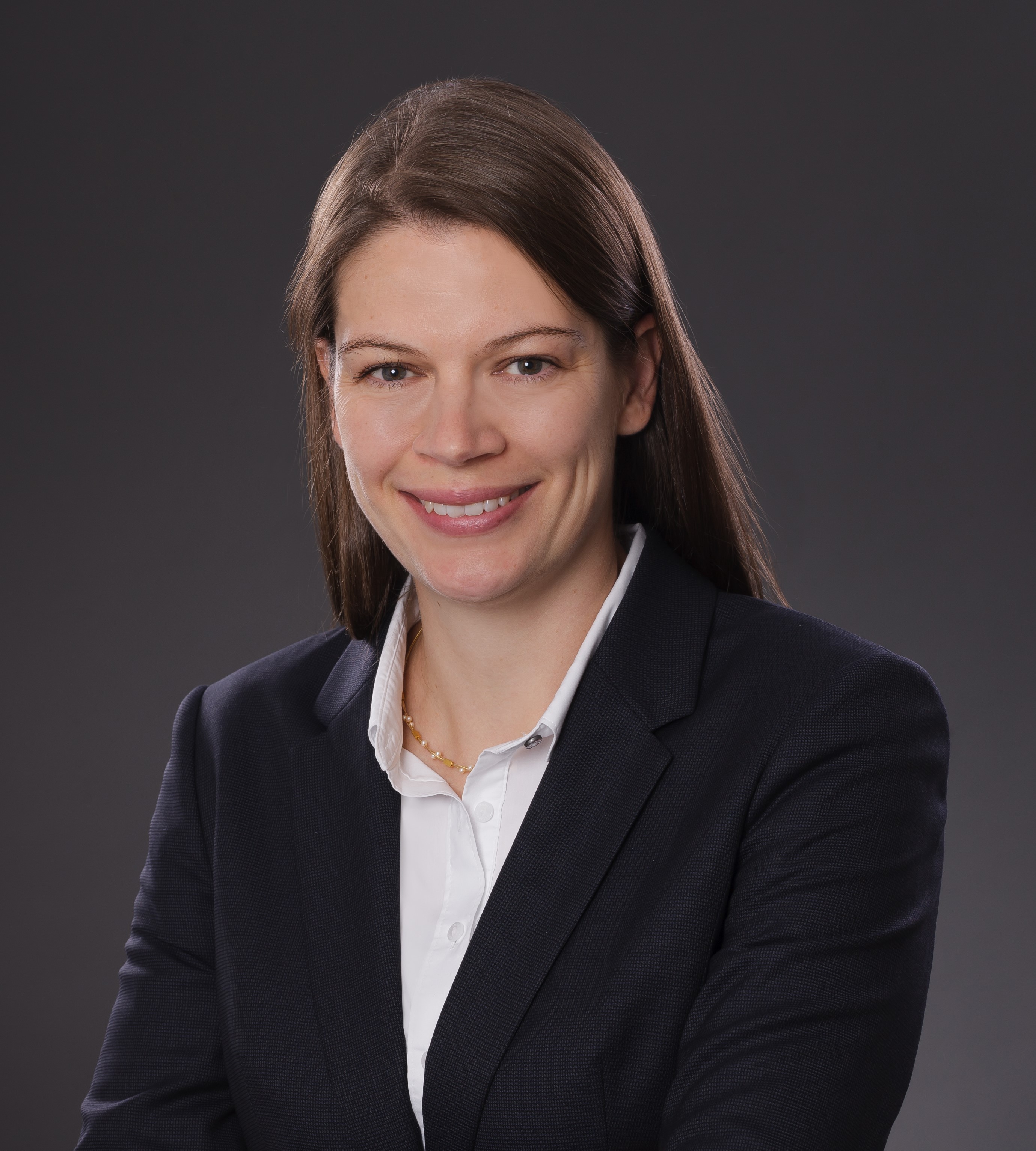}}]{Anke Schmeink}(IEEE Senior Member, Editor \emph{IEEE Trans. Wireless Commun.}, Prof. Dr.-Ing.) is leading the Chair of Information Theory and Data Analytics at RWTH Aachen University, Germany. She received the Diploma degree in mathematics with a minor in medicine and the Ph.D. degree in electrical engineering and information technology from RWTH Aachen University, Germany, in 2002 and 2006, respectively. She worked as a research scientist for Philips Research before joining RWTH Aachen University. She spent several research visits with the University of Melbourne, and with the University of York. She is co-author of more than 270 publications and editor of the books \emph{Big Data Analytics for Cyber-Physical Systems: Machine Learning for the Internet of Things} and \emph{Smart Transportation: AI Enabled Mobility and Autonomous Driving}. Her research interests are in information theory, machine learning, data analytics and optimization with focus on wireless communications and medical applications.
\end{IEEEbiography}
%\vfill

\begin{IEEEbiography}[{\includegraphics[width=1in,height=1.25in,clip,keepaspectratio]{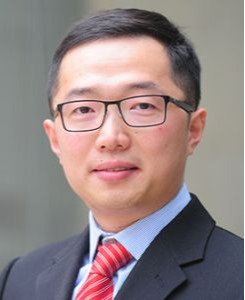}}]{Yan-Fu Li} (Senior Member, IEEE) was a Faculty Member with the Laboratory of Industrial Engineering, CentraleSup\'elec, University of Paris-Saclay, Gif-sur-Yvette, France, from 2011 to 2016. He is currently a Professor with the Industrial Engineering Department, Tsinghua University, Beijing, China. He has led/participated in several projects supported by the European Union (EU), France, and Chinese Governmental funding agencies and various industrial partners. He has coauthored or coauthored more than 100 publications on international journals, conference proceedings, and books. His current research interests include reliability, availability, maintainability, and safety (RAMS) assessment and optimization with the applications onto various industrial systems. Dr. Li is an Associate Editor of \emph{IEEE Transactions on Reliability}.
\end{IEEEbiography}
%\vfill

\begin{IEEEbiography}[{\includegraphics[width=1in,height=1.25in,clip,keepaspectratio]{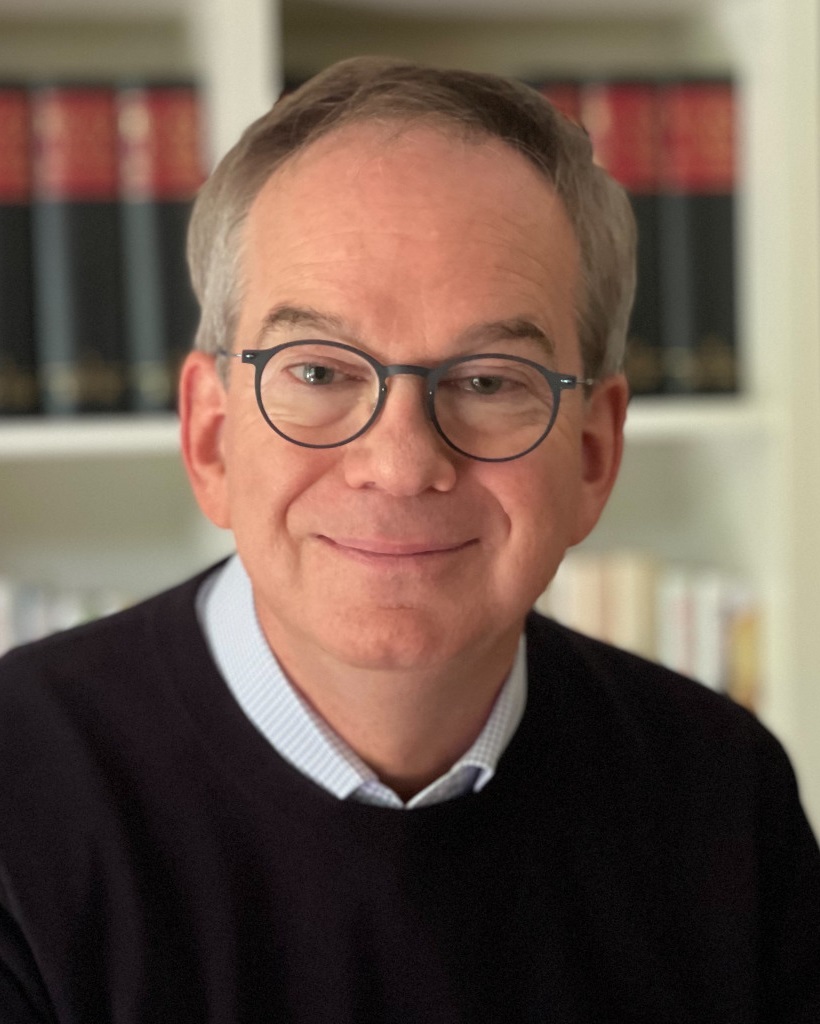}}]{Hans D. Schotten} (Member, IEEE) received the Ph.D. degree from the RWTH Aachen University, Germany, in 1997. From 1999 to 2003, he worked for Ericsson. From 2003 to 2007, he worked for Qualcomm. He became manager of a R\&D group, Research Coordinator for Qualcomm Europe, and Director for Technical Standards. In 2007, he accepted the offer to become a Full Professor at Technische Universit\"at Kaiserslautern. In 2012, he - in addition - became the Scientific Director of the German Research Center for Artificial Intelligence (DFKI) and Head of Department for Intelligent Networks. Professor Schotten served as Dean of Department of Electrical and Computing Engineering at Technische Universit\"at Kaiserslautern from 2013 until 2017. Since 2018, he is the Chair of the German Society for Information Technology and a member of the Supervisory Board of the VDE. He has authored more than 200 papers and participated in over 30 European and national collaborative research projects.
\end{IEEEbiography}

%\vfill

\appendices
\section{Proof of Lemma~\ref{lemma:one-shot}}\label{app:proof_one-shot}
\begin{proof}
	For the last hop ($i_s=1$),  consider an \emph{arbitrary} \ac{DMH-HARQ} scheme with maximal $K$ transmission attempts, where the blocklength of $k\superscript{th}$ (re-)transmission is $m_k$ for all  $k\in\{1,2,\dots K\}$. Obviously, since $i_s=1$, the utility equals the overall chance of successful transmission. Using Type I \ac{HARQ} without combining, that is
	\begin{equation}
		\begin{split}
			\xi=&1-\epsilon_I(m_1)+\epsilon_I(m_1)\left[1-\epsilon_I(m_2)\right]\\
			&+\epsilon_I(m_1)\epsilon_I(m_2)\left[1-\epsilon_I(m_3)\right]+\dots\\
			&+\epsilon_I(m_1)\dots\epsilon_I(m_{K-1})\left[1-\epsilon_I(m_K)\right]\\
			=&1-\prod\limits_{k=1}^{K}\epsilon_I(m_k)
		\end{split}
	\end{equation}
	Meanwhile, according to \cite{makki2014finite}, if it is Type-II \ac{HARQ} with \ac{IR} applied, the overall chance of successful transmission is equivalent to that of an \emph{one shot} transmission:
	\begin{equation}
		\xi\subscript{os}=1-\epsilon_I\left(\sum\limits_{k=1}^Km_k\right).
	\end{equation}
	Obviously, under the same blocklength allocation and channel conditions, Type II \ac{HARQ}-\ac{IR} always outperforms Type I \ac{HARQ} without combining in error rate, i.e., $\prod\limits_{k=1}^K\epsilon_I(m_k)\geqslant \epsilon_I\left(\sum\limits_{k=1}^Km_k\right)$, which takes the equality only when $K=1$. Furthermore, given the fixed remaining time $t_s$, we always have the constraint
%	\begin{equation}
		$\sum\limits_{k=1}^K\left(m_kT\subscript{b}+T\subscript{l}\right)\leqslant t_s$,
%	\end{equation}
	which means 
	\begin{equation}
		\sum\limits_{k=1}^Km_k\leqslant \frac{t_s-KT\subscript{l}}{T\subscript{b}}\leqslant\frac{t_s-T\subscript{l}}{T\subscript{b}},
	\end{equation}
	where the first equality is taken only when $t_s$ is sufficiently exploited, and the second taken only when $K=1$. 
	Additionally, i always holds that $\partial\epsilon_I(m)/\partial m<0$ for all $m\in\mathbb{R}^+$. Thus, we have 
	\begin{equation}
		\prod\limits_{k=1}^K\epsilon_I(m)\geqslant \epsilon_I\left(\frac{t_s-T\subscript{l}}{T\subscript{b}}\right),
	\end{equation}
	where the equality is only taken when $K=1$ and $m_1=\frac{t_s-T\subscript{l}}{T\subscript{b}}$, and the Lemma is therewith proven.
	
	%	to hold that $\partial\epsilon_I(n)/\partial n<0$ for all $n\in\mathbb{R}^+$, and $\partial^2\epsilon_I(n)/\partial n^2>0$ for all $n\in\mathbb{R}^+$
\end{proof}

\section{Proof of Theorem~\ref{th:continuous_monotone_utility}}\label{app:proof_continuous_monotone}
\begin{proof}
	For the first step, we prove the monotone of $\xi\subscript{max}(s)$ regarding $t_s$ as follows. Given any feasible system status $s_0=(t_{s_0}, i_{s_0}, k_{s_0}, \tau_{s_0})$, let $n\subscript{opt}$ be an optimal blocklength allocation policy\footnote{The existence of such optima is guaranteed by the bounded range of $\xi(s_0)$. However, the uniqueness of $n\subscript{opt}$ is not guaranteed. Hence we use the term ``an optimal policy'' instead of ``the optimal policy'' here.} that generates the DMH-HARQ schedule $\mathfrak{T}_{s_0}\superscript{opt}$ with maximized utility $\xi\subscript{max}(s_0)$. According to Lemma~\ref{lemma:one-shot}, there must be $n\subscript{opt}(s)=\frac{t_s}{T\subscript{b}}-T\subscript{l}$ for all $s$ that $i_s=1$.
	
	Given an \emph{arbitrary} $\Delta t>0$, let $s^{<\Delta t>}$ denote $(t_s+\Delta t, i_s, k_s, \tau_s)$, we can always construct a sub-optimal $n\subscript{sub}$ that
	\begin{equation}
		n\subscript{sub}(s)
		=\begin{cases}
			n\subscript{opt}(s^{<-\Delta t>})+\frac{\Delta t}{T\subscript{b}}&i_s=1,\\
			n\subscript{opt}(s^{<-\Delta t>})&\mathrm{otherwise}.
		\end{cases}
	\end{equation}
	Now apply $n\subscript{sub}$ on the initial state $s_1=s_0^{\Delta t}$ to generate another \ac{DMH-HARQ} schedule $\mathfrak{T}\superscript{sub}_{s_1}$, it is straightforward to derive that $\mathfrak{T}\superscript{opt}_{s_0}$ and $\mathfrak{T}\superscript{opt}_{s_1}$ have the identical structure, except that for each non-empty node $s$ in $\mathfrak{T}\superscript{opt}_{s_0}$, the corresponding node of $\mathfrak{T}\superscript{opt}_{s_1}$ is 
	\begin{equation}
		s'=\begin{cases}
			s&i_{s'}=0,\\
			s^{<\Delta t>}&\text{otherwise}.
		\end{cases}
	\end{equation}
	And error rates of each transmission attempt regarding the two schedules, denoted as $\varepsilon_s$ and $\varepsilon'_{s'}$, respectively, must fulfill
	\begin{equation}
		\varepsilon'_{s'}\begin{cases}
			<\varepsilon_s&i_{s'}=1,\\
			=\varepsilon_s&\text{otherwise}.
		\end{cases}
	\end{equation}
	Recalling Eq.~\eqref{eq:bellman}, it always hold $\xi(\mathfrak{T}\superscript{sub}_{s_1})>\xi\subscript{max}(s_0)$. Meanwhile, there is always some $n'\subscript{opt}$ that generates an optimal schedule $\mathfrak{T}\superscript{opt}_{s_1}$ with maximal $\xi\subscript{max}(s_1)\geqslant \xi(\mathfrak{T}\superscript{sub}_{s_1})$. Thus:
	\begin{equation}
		\xi\subscript{max}(s_1)=\xi\subscript{max}(s_0^{<\Delta t>})>\xi\subscript{max}(s_0)\quad\forall s_0, \forall\Delta t>0,
	\end{equation}
	i.e. $\xi_{\mathrm{max}}(s)$ is strictly monotone increasing regarding $t_s$.
	
	Next, we come to prove the continuity of this function. According to Eq.~\eqref{eq:err_model_harq_1_wo_sc}, $\varepsilon_s$ is continuous function of $t_s$, so $\xi\subscript{max}(s)$ can be guaranteed $t_s$-continuous, if both $\xi\subscript{max}(s\subscript{l}(\mathfrak{T}_s))$ and $\xi\subscript{max}(s\subscript{r}(\mathfrak{T}_s))$ are continuous functions of $t_s$. This sufficient condition can be derived via mathematical induction as follows. 
	
	Starting with the two \emph{termination cases} of $t_s<T\subscript{min}^{(i_s)}$ and $i_s=0$, respectively, according to Eqs.~\eqref{eq:left_branch_binary_tree} and \eqref{eq:right_branch_binary_tree} we have $S(L(\mathfrak{T}_s))=S(R(\mathfrak{T}_s))=\emptyset$. Recalling \eqref{eq:recursive_utility}, for the termination cases we have $\xi\subscript{max}(s)=\xi(\mathfrak{T}_s)\equiv 0$, which are both $t_s$-continuous. Therefore, the utility $\xi\subscript{max}(s)$ itself is also $t_s$-continuous.
	
	Then consider the last hop where $i_s=1$, according to Lemma~\ref{lemma:one-shot} we know that $\xi\subscript{max}(s)$ is achieved when $n_s=\frac{t_s-T\subscript{l}}{T\subscript{b}}$, leading to a left branch where $t_{s\subscript{l}(\mathfrak{T}_s)}=0<T\subscript{min}^{(1)}$ and a right branch where $i_{s\subscript{r}(\mathfrak{T}_s)}=0$, both are termination cases. Thus, as proven above we have both $\xi\subscript{max}(s\subscript{l}(\mathfrak{T}_s))$ and $\xi\subscript{max}(s\subscript{r}(\mathfrak{T}_s))$ are $t_s$-continuous, and therefore $\xi\subscript{max}(s)$ the same as well.
	
	For other hops other than the last one, i.e. $i_s\in\{2,3,\dots I\}$, according to the DMH-HARQ principle \eqref{eq:left_branch_binary_tree} and \eqref{eq:right_branch_binary_tree}, it is obvious that there are only three possible combinations of its left and right branches:
	\begin{enumerate}
		\item $s\subscript{l}(\mathfrak{T}(s))=\emptyset, i_{s\subscript{r}(\mathfrak{T}(s))}=\emptyset$;
		\item $s\subscript{l}(\mathfrak{T}(s))=\emptyset, i_{s\subscript{r}(\mathfrak{T}(s))}=i_s-1$;
		\item $i_{s\subscript{l}(\mathfrak{T}(s))}=i_s, i_{s\subscript{r}(\mathfrak{T}(s))}=i_s-1$.
	\end{enumerate}
	For case 1), $\xi\subscript{max}(s)\equiv 0$, which is $t_s$-continuous. For case 2), $\xi\subscript{max}(s\subscript{l}(\mathfrak{T}_s))$ is $t_s$-continuous as proven above, so that $\xi\subscript{max}(s)$ can be guaranteed $t_s$-continuous if only $\xi\subscript{max}(s\subscript{r}(\mathfrak{T}_s))$ is $t_s$-continuous. For case 3), we recall Lemma~\ref{lemma:history_independence} that $n\subscript{opt}(s)$ depends only on $t_s$ and $i_s$, so that $\xi\subscript{max}(\mathfrak{T}(s))$ and $\xi\subscript{max}(L(\mathfrak{T}(s)))$ are different samples for the \emph{same} function of $t_s$. Since the iterative generation of the left branch in decision tree is determined to terminate with an empty set, i.e. case 1) or 2), we can conclude that for a $\mathfrak{T}s$ in case 3), both $\xi\subscript{max}(s)$ and $\xi\subscript{max}(s\subscript{l}(\mathfrak{T}(s)))$ are guaranteed $t_s$-continuous if $\xi\subscript{max}(s\subscript{r}(\mathfrak{T}_s))$ is $t_s$-continuous. Since the iterative generation of the right branch in decision tree always ends up with one of the two termination cases, which are both $t_s$-continuous, we can conclude from the analyses above that $\xi_{\mathrm{max}}(s)$ is also $t_s$-continuous for all $s$ where $i_s>1$. 
	
	Thus, the continuity is proven for all $i\in\mathcal{S}$. With both the monotone and the continuity derived, the theorem is proven.
\end{proof}

\end{document}